\documentclass{emulateapj}

\usepackage{amsmath,natbib, float, color}
\newcommand*\Msolarh[0]{\mathrm{M_{\odot}}\, h^{-1}}
\newcommand*\Mpch[0]{\mathrm{Mpc} \, h^{-1}}

\newcommand*\vlos[0]{v_{\mathrm{los}}}

\newcommand*\kms[0]{\mathrm{km \, s^{-1}}}

\newcommand*\Mtot[0]{M_\mathrm{true}}
\newcommand*\Mdyn[0]{M_\mathrm{dyn}}
\newcommand*\vtrue[0]{v_\mathrm{true}}
\newcommand*\verr[0]{v_\mathrm{err}}
\newcommand*\model[0]{$\sigma_8$-$\Omega_m$}
\newcommand*\mathmodel[0]{\sigma_8, \, \Omega_m}

\newcommand*\sig[0]{$\sigma_8$}
\newcommand*\OM[0]{$\Omega_m$}

\newcommand*\new[1]{{#1}}

\newcommand*\updated[1]{#1}

\shorttitle{Velocity Distribution Function}
\shortauthors{Ntampaka et al.}

\begin{document}

\title{The Velocity Distribution Function of Galaxy Clusters  as a Cosmological Probe}

\author{M. Ntampaka\altaffilmark{1}, H. Trac\altaffilmark{1}, J. Cisewski\altaffilmark{2}, and L.C. Price\altaffilmark{1}}
\email{ntampaka@cmu.edu}
\altaffiltext{1}{McWilliams Center for Cosmology, Department of Physics, Carnegie Mellon University, Pittsburgh, PA 15213}
\altaffiltext{2}{Department of Statistics, Yale University,  New Haven, CT 06520}
\affil{}

\begin{abstract}
We present a new approach for quantifying the abundance of galaxy clusters and constraining cosmological parameters using dynamical measurements. In the standard method, galaxy line-of-sight velocities, $v$, or velocity dispersions are used to infer cluster masses, $M$, to quantify the halo mass function (HMF), $dn(M)/d\log(M)$, which is strongly affected by mass measurement errors. In our new method, the probability distribution of velocities for each cluster in the sample are summed to create a new statistic called the velocity distribution function (VDF), $dn(v)/dv$. The VDF can be measured more directly and precisely than the HMF and can be robustly predicted with cosmological simulations which capture the dynamics of subhalos or galaxies. We apply these two methods to realistic (ideal) mock cluster catalogs with (without) interlopers and forecast the bias and constraints on the matter density parameter $\Omega_m$ and the amplitude of matter fluctuations $\sigma_8$ in flat $\Lambda$CDM cosmologies. For an example observation of 200 massive clusters, the VDF with (without) interloping galaxies constrains the parameter combination 
$\smash{\sigma_8\,\Omega_m^{0.29\ (0.29)}} =  0.589 \pm 0.014 \ (0.584 \pm 0.011)$ 
and shows only minor bias. However, the HMF with interlopers is biased to low $\Omega_m$ and high $\sigma_8$ and the fiducial model lies well outside of the forecast constraints, prior to accounting for Eddington bias. When the VDF is combined with constraints from the cosmic microwave background, the degeneracy between cosmological parameters can be significantly reduced. Upcoming spectroscopic surveys that probe larger volumes and fainter magnitudes will provide clusters for applying the VDF as a cosmological probe.
\end{abstract}


\section{Introduction}
\label{sec:intro}

Galaxy clusters are used to probe the fundamentals of cosmology, from the growth of structure to parameters describing the underlying cosmological model.  Yet, measurement errors plague these techniques; they are the primary source of bias and uncertainty in applying clusters as cosmological probes. Galaxy clusters are massive, rare objects that populate the high-mass tail of the halo mass function (HMF), and their abundance as a function of mass and redshift depends sensitively on the underlying cosmological model (See \cite{2005RvMP...77..207V, 2011ARA&A..49..409A} for a review).  In a flat $\Lambda$CDM cosmologies, massive cluster abundance is particularly sensitive to the matter density parameter $\Omega_m$ and the amplitude of matter fluctuations $\sigma_8$ {, and even observations of a few very massive clusters can rule out cosmological models \citep[e.g.][]{1998ApJ...504....1B}.}  However, measurement errors, in combination with the steeply-declining HMF, introduce biases that limit the reliability of the HMF as a cosmological probe.

A variety of techniques exist for determining cluster masses from observables, employing observations in multiple wavelengths.  For example, X-ray observations of hot intracluster medium (ICM) can be used to infer a cluster mass profile \citep[e.g.][]{2009ApJ...692.1033V, 2010MNRAS.406.1773M}. \citet[SZ;][]{1972CoASP...4..173S} distortions of the cosmic microwave background (CMB) by the hot ICM can be used to determine a temperature-weighted gas mass \citep[e.g.][]{2010ApJ...719.1045L, 2013JCAP...07..008H}. Strong and weak gravitational lensing of distant sources directly probes total mass along the line of sight (LOS), and can be used to determine projected mass profiles \citep[e.g.][]{2007arXiv0709.1159J, 2008JCAP...08..006M}.  {The virial theorem may be applied to relate LOS velocity dispersion, $\sigma_v$, to cluster mass as a power law \citep[e.g.][]{1990ApJS...72..715T, 2014ApJ...792...45R}.  The caustic technique \new{\citep{1997ApJ...481..633D, 1999MNRAS.309..610D}} uses both LOS velocities and projected distances to probe a cluster's mass distribution \citep[e.g.][]{1999MNRAS.309..610D, 2007ApJ...657..183R}, showing good agreement with x-ray mass profiles \citep{2015arXiv151107872M} and also with weak lensing masses in well-studied systems \citep[e.g.][]{2014ApJ...783...52G}.  Further, these mass profiles can be used to probe the internal structure of clusters including mass-concentration relations and density profiles \citep[e.g.][]{2006AJ....132.1275R, 2013ApJ...767...15R, 2015ApJ...806....4M}.  }

Yet, sources of error in each of these methods limit the accuracy with which halo masses can be predicted. X-ray analyses assume hydrostatic equilibrium, but biases can arise because of nonthermal pressure support \citep[e.g.][]{1990ApJ...363..349E, 2004MNRAS.351..237R, 2009ApJ...705.1129L}. SZ scaling relations derived using X-ray measurements \citep[e.g.][]{2010A&A...517A..92A} can also have hydrostatic mass bias, while those calibrated using simulations are affected by uncertainties in ICM astrophysics \citep[e.g.][]{2006ApJ...650..538N, 2012ApJ...758...74B}. Scatter in weak lensing masses can be caused by matter along the line of sight and halo triaxiality \citep[e.g.][]{2001A&A...370..743H, 2011ApJ...740...25B}.  Dynamical cluster mass sources of scatter include halo environment and triaxiality \citep[e.g.][]{White:2010ab, 2012MNRAS.426.1829N, 2014arXiv1405.0284S} and mass and galaxy selection strategy \citep{2013MNRAS.434.2606O, 2013ApJ...772...47S}.   {The caustic technique reduces scatter compared to a velocity dispersion scaling relation in some cases \citep{2013ApJ...773..116G}, but has comparable scatter in others \citep[e.g.][]{2014MNRAS.441.1513O}.  Attempts have been made to correct for these sources of error \citep[e.g.][]{2014MNRAS.441.3562E, 2011PhRvD..83b3015M}, though the corrections may come at the expense of reduced constraining power}

This scatter in the mass-observable relationship alters the observed distribution of cluster masses from the true distribution, an effect known as the Eddington bias \citep{1913MNRAS..73..359E}.  Scatter in predicted mass, when combined with a steeply-declining HMF that spans several orders of magnitude, results in an observed HMF that differs from the true HMF.  This is most notable at the high-mass end.  Caused by abundant, lower-mass halos scattering into higher mass bins due to measurement error in their masses, the resulting shape and amplitude of the observed HMF shows an upscatter at high masses.  When the Eddington bias is uncorrected, the result is a bias in the maximum likelihood \sig{} and \OM{} predicted by an HMF analysis of cluster observations.

 {Rather than quantifying cluster abundance via a HMF, it can rather be quantified with the distribution of a summary statistic, for example, velocity dispersion $\sigma_v$. Velocity dispersion counts can be used to constrain cosmological parameters \citep{Evrard:2008aa, 2016arXiv160200611C}, and it has been noted that the high $\sigma_v$ tail of this distribution provides a constraint on scenarios of structure formation \citep{1996A&A...310...31M}.  The distribution of velocity dispersions has been used to probe the bias between the velocity dispersions of observed cluster galaxies and simulated cluster dark matter  \citep{2007ApJ...657..183R}.} 

{Cluster abundance can also be described by} the distribution of direct observables to minimize the biases. For example, \cite{2012PhRvD..86l2005W} and \cite{2014arXiv1411.8004H} have directly analyzed the probability distribution function (PDF) of thermal SZ temperature fluctuations rather than cluster counts. This statistic has recently been measured to high precision \citep{2015arXiv150201596P}, but there are current limitations in using it to constrain cosmological parameters. This approach does not require converting the integrated Compton-$y$ flux to mass using a scaling relation, but it does require knowledge of ICM pressure profiles as a function of halo mass to construct templates for comparison.

In this paper, we present a new approach for quantifying the abundance of galaxy clusters using dynamical measurements. Galaxy LOS velocities, $v$, can be accurately measured with spectroscopic surveys or followups and they can also be robustly predicted with cosmological simulations which capture the gravitationally-induced dynamics of subhalos or galaxies. To include more observable information, a given cluster is represented by the probability distribution of galaxy velocities rather than a single velocity dispersion. We construct a new statistic called the velocity distribution function (VDF) by summing the PDF($v$) for $N$ most-massive or most-richest clusters in a known volume $V$. Clusters can also be ranked relatively using other mass proxies, but absolute mass measurements are not necessary.

We show that the VDF can be measured more directly and precisely than the HMF. For the latter, we follow the conventional approach of using velocity dispersion to infer dynamical mass and do not attempt to correct for scatter in dynamical mass and the resulting Eddington bias. Section 2 describes how the HMF and VDF are constructed using a large mock cluster catalog. Section 3 compares the constraining power of these two statistics on \OM{} and \sig{}. The implications are discussed in Section \ref{sec:discussion} and the conclusions are summarized in Section \ref{sec:conclusion}.

\section{Methods}
\label{sec:methods}

\subsection{Mock Catalog}

This work is based on mock cluster observations created from the publicly available Multidark MDPL1 $N$-body simulation\footnote{http://www.cosmosim.org/}.  The simulation contains $3840^3$ particles in a box of length $1\ h^{-1}\rm{Gpc}$ with mass resolution of  $1.51\times10^9\, \Msolarh$.  Multidark employs a $\Lambda$CDM cosmology, with cosmological parameters consistent with Planck data \citep{2014A&A...571A..16P}:  $\Omega_{\Lambda} = 0.693$, $\Omega_m = 0.307$, $\Omega_b = 0.048$, $h = 0.678$, $n=0.96$, and $\sigma_8 = 0.823$.

Halos for the mock observations are chosen from the publicly available BDMW halo catalog. This  halo catalog employs a bound density maximum (BDM) spherical overdensity halo finder with halo average density equal to 200 times the critical density of the Universe \citep{1997astro.ph.12217K}.  The $z=0$ BDMW halo catalog is chosen for this work due to the large number of massive halos at low redshift.  A lower mass cut is imposed on halos; only those with $M \geq 10^{14} \ \mathrm{M_{\odot}} h^{-1}$ are included.  

In addition to the halo mass and redshift requirement, a richness cut is imposed such that clusters in the mock catalog must contain at least $20$ BDMW subhalos with $M_{\mathrm{subhalo}} \geq 10^{12} \ \mathrm{M_{\odot}}h^{-1}$.  Halos and their subhalo structure are presumed to represent a galaxy, with the galaxy inheriting its host halo or subhalo's position and velocity.  

\begin{deluxetable*}{l l r l r r r r r}

\tablecaption{Catalog Summary\label{table:catalog}} 
\tablewidth{0pc} 
\tablehead{ 
\colhead{Name} 
&\colhead{Cluster} 
& \colhead{Velocity Error}
& \colhead{Mass}
&\colhead{$\alpha$\tablenotemark{a}}  
& \colhead{$\sigma_\mathrm{15}$\tablenotemark{a}}     
& \colhead{Aperture}  
& \colhead{Velocity cut} 
& \colhead{Unique} \\
&\colhead{Membership} 
& \colhead{$(\kms)$}
& \colhead{}
&\colhead{}  
& \colhead{$(\kms)$}     
& \colhead{$(\Mpch)$}  
& \colhead{$(\kms)$} 
& \colhead{Clusters\tablenotemark{b}}
}

\startdata 

\new{Idealized}				& pure and complete				& 0		& $\Mtot{}$		& \nodata 		&  \nodata  	&\nodata 	& \nodata 			& \new{1909}   \\[1ex]
Intermediate		& pure and complete				& $100$ 	& $\Mdyn{}$		&$0.37$ 		& 1254 		&\nodata	& \nodata	& \new{1909}   \\[1ex]
Contaminated		& cylindrical cut with 2-$\sigma$ clip		& $100$	& $\Mdyn{}$		& $0.32$ 		& $ 1124$ 		&$1.6$ 	& $2500$	& \new{1910}  \\

\enddata

\tablenotetext{a}{Fit parameters to Equation \ref{eq:powerlaw}. }
\tablenotetext{b}{With $\Mtot{}\geq\ 3.5\times10^{14}\,\Msolarh$ and at least 20 member or interloping galaxies.}

\end{deluxetable*} 

{We create three mock catalogs:
\begin{enumerate}
\item{} \new{Idealized}:  This mock catalog is ideal.  It has pure and complete cluster membership information and perfect mock measurements of LOS velocities.  The true dark matter halo mass, $\Mtot{}$, is presumed to be known.  Further details of this catalog can be found in \cite{Ntampaka2015}. 
\item{} Intermediate:  Like the \new{Idealized} catalog, this mock catalog has pure and complete cluster membership information.  However, realistic errors are added to LOS velocities and dynamical mass estimates from these velocities, $\Mdyn{}$, replace $\Mtot{}$ as the observed mass.    
\item{} Contaminated:  Realistic cluster mock observations should account for interloping galaxies that appear to belong within the cluster but actually reside in the cluster's fore- or background.  Thus, for this catalog, cylindrical cuts are made around the cluster centers, allowing interloping galaxies to contaminate the clusters.  All cylinders have a fixed aperture of $1.6\,\Mpch$ and initial velocity cut of $2500\,\kms$.  \new{The aperture and velocity cuts are chosen to contain a typical $M=10^{15}\,\Msolarh$ halo.}  Velocity outliers beyond 2$\sigma$ are iteratively removed from the sample until convergence.  Further details of this method can be found in \cite{Ntampaka2015b}
\end{enumerate}
 \new{For the Idealized and Intermediate Catalogs, richness refers to the number of members within $R_{200}$; for the Contaminated Catalog, richness refers to the number of objects remaining in the cylinder after velocity paring.}  These three mock catalogs provide a range of constraints from an ideal case (\new{Idealized}) to a more realistic catalog (Intermediate) as well as to show constraints in a catalog with interlopers (Contaminated).

\new{For the Idealized and Intermediate Catalogs, we have applied a straightforward spherical cut, choosing substructure lying within the halo $R_\mathrm{200}$.  Applying a method such as the shifting gapper or caustics method can create a sample that is nearly pure and complete.  The caustics method reports completeness $\approx 95\%$ and contamination $\approx 2-8\%$ \citep{2013ApJ...768..116S}.  For the Contaminated Catalog, the fixed aperture cylinder is a simple cut that is straightforward to apply observationally.  An adaptive cylindrical cut utilizing, for example, an iterative scheme \citep{2007MNRAS.379..867V} or cluster mass proxy could be used to more closely approximate the true $R_{200}$.  Applying such a method would reduce the impact of interlopers, resulting in a catalog that is more closely approximated by the Intermediate Catalog.}}

Mock catalog details are summarized in Table \ref{table:catalog} for reference.  \new{The number of unique clusters with $\Mtot{}\geq3.5 \times 10^{14}\,\Msolarh$ and at least 20 member or interloping galaxies is noted.  Note that, below this mass, the number of unique objects in the pure and Contaminated Catalogs deviate.  The \new{Idealized}  and Intermediate Catalogs fall below the true HMF due to the imposed richness cut.  The Contaminated Catalog, however, follows the true HMF more closely due to interloping galaxies allowing galaxy-poor clusters to appear to have richness above the minimum cut.}

We explore mock observations of several sizes {to explore how large future surveys might constrain cosmological parameters as well as to forecast cosmological constraints with available observations of $N\approx50$ clusters \citep[e.g.][]{2014ApJ...792...45R, 2015arXiv151200910S}.}  Cubical mock observations containing a mean of $N=200$ clusters with $M\geq3.5\times10^{14}\,\Msolarh$ are selected from the simulation box.  This choice of minimum mass and cluster number dictate a mock observation volume, with resulting side length $481.5\,\Mpch$.  A smaller observation denoted $N=100$ ($N=50$) has half (one-quarter) the average number of clusters in half (one-quarter) of the volume, with the same lower mass limit.  {Cosmological-volume simulations are computationally expensive, and there are few publicly available options.  To take full advantage of the simulation, it is necessary to use overlapping volumes; our mock observations overlap in a $8\times8\times8$ grid within the full simulation volume}.  Three LOS views of each of the $512$ mock observations {along the simulation $x$-, $y$-, and $z$-directions} are unique, albeit correlated, observations.

\subsection{Halo Mass Function}

\begin{figure}[!t]
\begin{center}
\begin{tabular}{c c}
        	\includegraphics[width=0.45\textwidth]{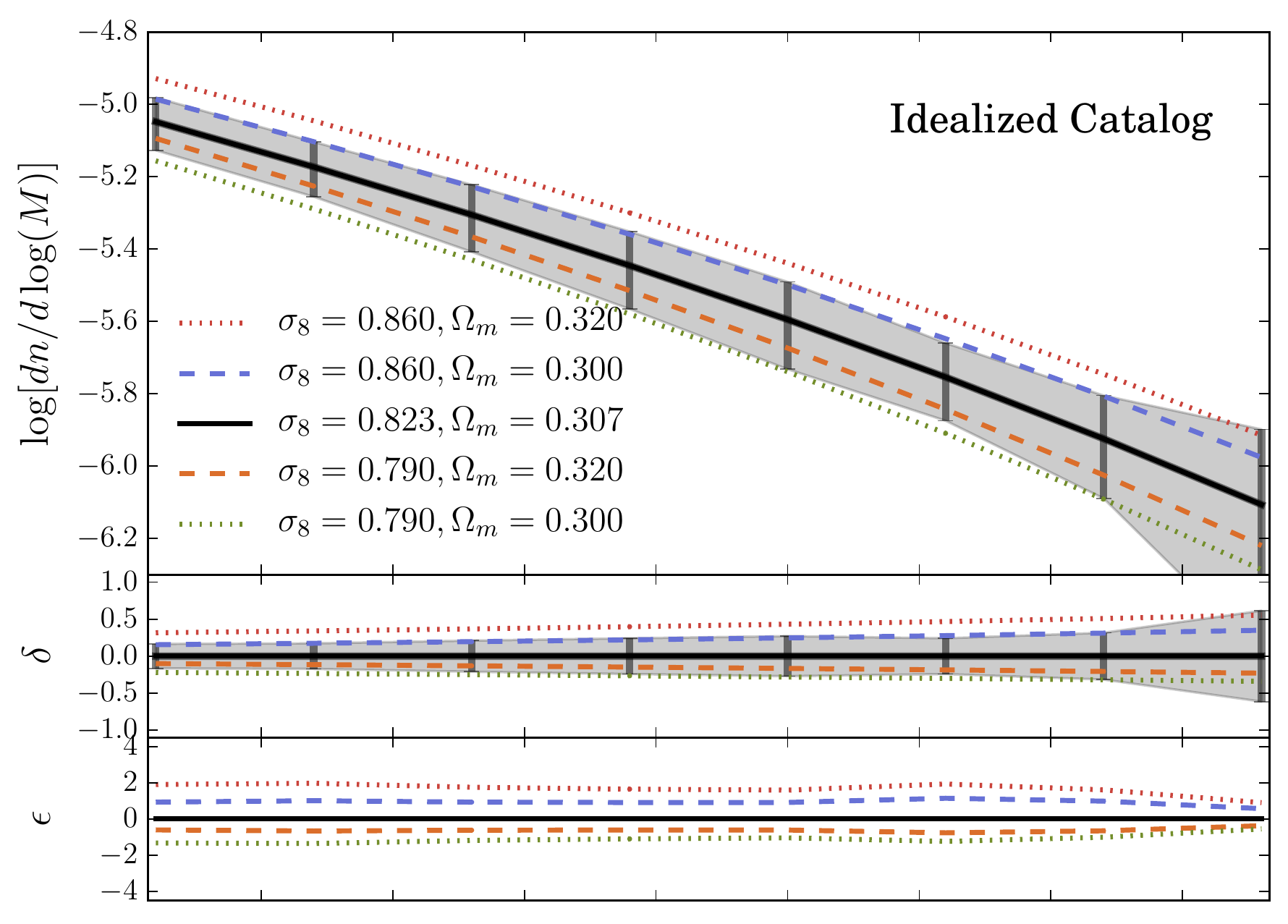} & \\
	\includegraphics[width=0.45\textwidth]{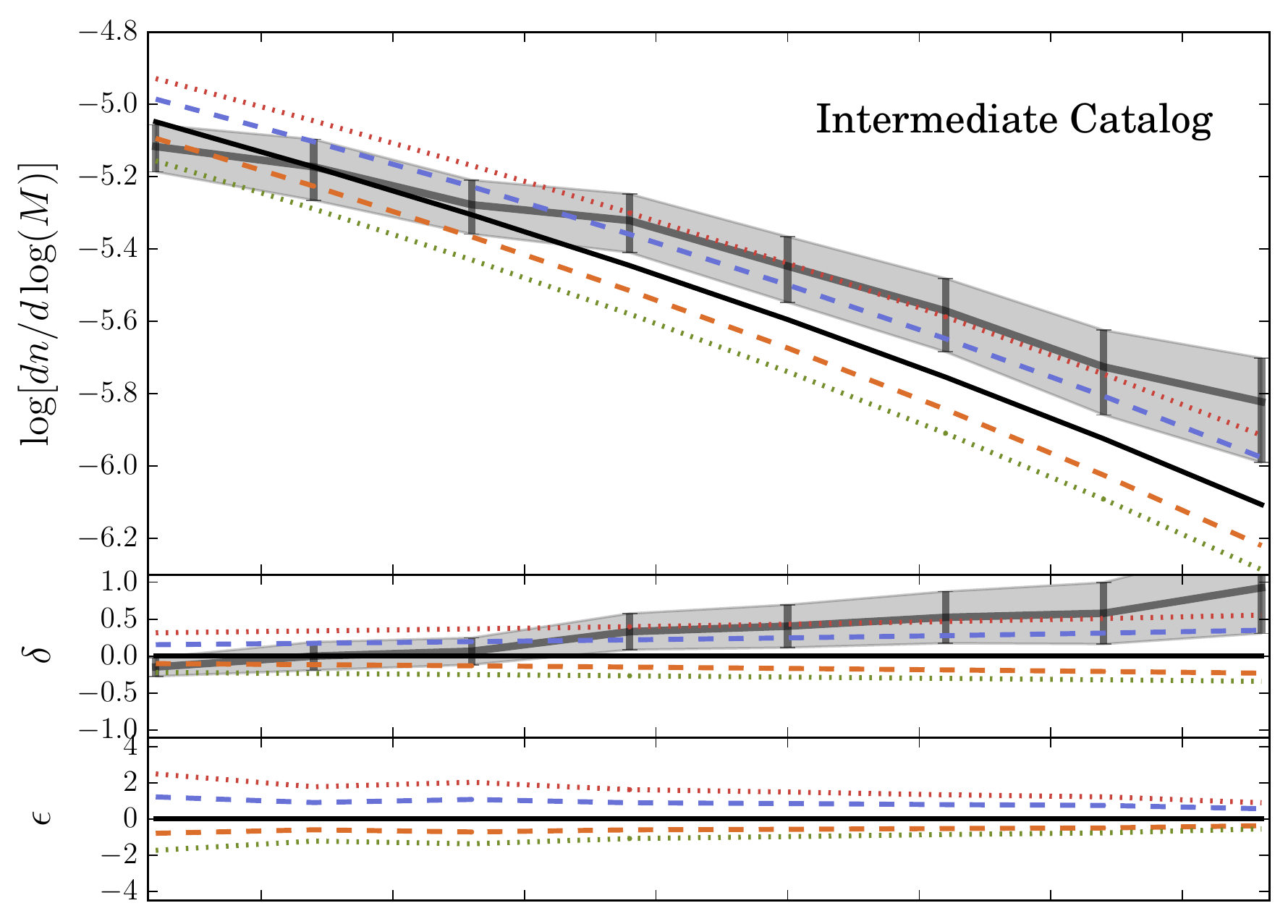} & \\
	\includegraphics[width=0.45\textwidth]{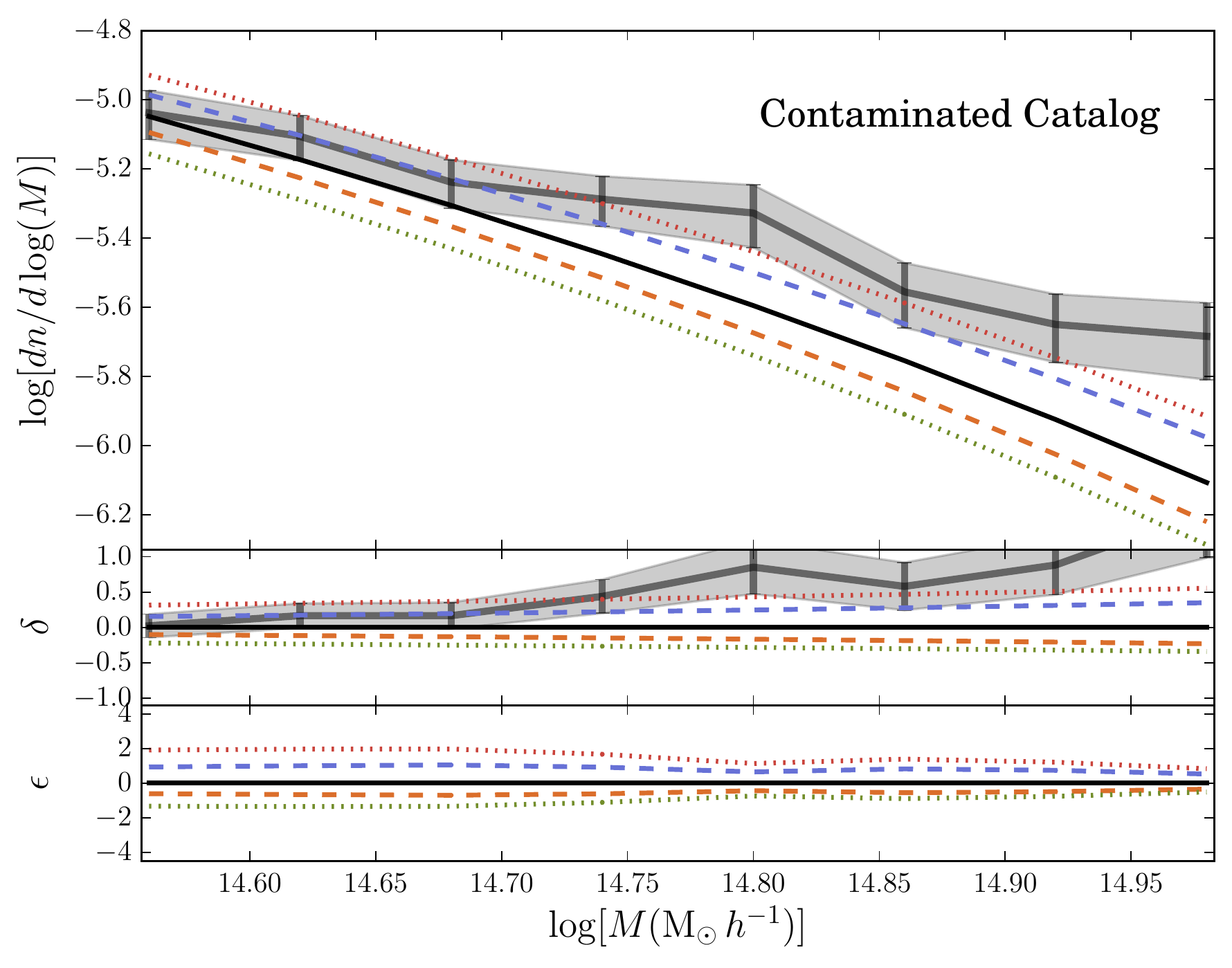} & \\
\end{tabular}
\end{center}

	\caption{\footnotesize{{{Top:  Halo mass function for massive clusters assuming {the \new{Idealized} Catalog with} true masses, $\Mtot{}$. For the fiducial model (black solid), the expected 68\% scatter {(light gray band)} is shown for mock observations with $\approx200$ clusters with mass $\Mtot{}\geq3.5\times\,10^{14}\,\Msolarh$.  Four additional \model{} model predictions (colored dashed and dotted) are shown for comparison (top).  The fractional difference between models, $\delta$, increases with mass, as does the scatter in mock observations (middle), while the normalized difference, $\epsilon$, decreases with mass, showing that lower-mass bins have the most resolving power (bottom). 
	{Middle}:  HMF constructed from the Intermediate Catalog clusters with clusters with $\Mdyn{}\geq3.5\times\,10^{14}\,\Msolarh$.  In contrast with the \new{Idealized} Catalog HMF, this catalog overpredicts the number of high mass clusters, and the mean (dark gray) does not agree with the fiducial model.
	Bottom:  the overprediction of high mass clusters is particularly pronounced in the Contaminated Catalog. }}
	}}
	\label{fig:hmf}
\end{figure}

Because cluster abundance is sensitive to the underlying cosmology, these massive objects can be utilized to constrain cosmological models.  The HMF constructed from the mock catalog is used here to forecast constraints on \sig{} and \OM{} with cluster counts as a function of mass.  Using the simulated clusters, we construct a halo mass function, $dn/d\log(M)$, from the true cluster masses.  This HMF will be compared to the HMF predicted by various \model{} models assuming a flat $\Lambda$CDM cosmology.    

We use clusters with {true mass $\Mtot{}\geq3.5\times10^{14}\,\Msolarh$ in the analysis of the \new{Idealized} catalog HMF};  below this mass, the richness cut imposed on the catalog creates a shortage of clusters.  The HMF's mass bin sizes are chosen by Freedman-Diaconis' rule \citep{Freedman:1981aa}, which relates bin width, $dx$, to the interquartile range of the data to be binned, $\mathrm{IQR}(\{x\})$, and the number of items to be binned, $N$,
\begin{equation}
dx = 2 \frac{\mathrm{IQR}(\{x\})}{N^{1/3}}.
\label{eq:FD}
\end{equation}
This binning rule of thumb is selected because it is less sensitive to outlying data and does not assume a normal distribution of data.  It is applied with $\{x\}=\{\log(M)\}$, and the upper mass limit is chosen where the ratio of mock observation mean halo counts to bin dispersion goes below 3.  The result is eight bins with $\Delta\log[M(\Msolarh)]=0.06$ and $14.53\leq\log(M)\leq15.01$.

Figure \ref{fig:hmf} shows the Multidark HMF, $dn/d\log(M)$, defined as the differential comoving number density, $n$ ($h^3\,\mathrm{Mpc}^{-3}$), of halos per unit mass, $M$ ($\Msolarh$).  {To create the \model{} model predictions for the HMF, we first employ the analytic mass function given by \cite{2008ApJ...688..709T}.} Because the Multidark simulation disagrees slightly with this model, each bin of the simulation HMF is scaled to match the fiducial analytic HMF.  This results in the simulation and fiducial analytic model agreeing exactly by construction, evident in Figure \ref{fig:hmf}.

Also shown in Figure \ref{fig:hmf} are several representative \model{} model mass functions, chosen to vary from the fiducial case by $\approx5\%$ in both \sig{} and \OM{}.  As \sig{} and \OM{} vary from the fiducial case, the model predictions disagree with the measured mass function.  In order to compare the HMF from different models, we define a fractional difference 
\begin{equation}
	\delta = \frac{y_\mathrm{model}-y_\mathrm{fiducial}}{y_\mathrm{fiducial}}, 
	\label{eq:epsilon}
\end{equation}
where $y_\mathrm{model}$ is a non-fiducial \model{} HMF and $y_\mathrm{fiducial}$ is the fiducial HMF.  While the highest mass bins have the largest fractional difference $\delta$, they also have the highest variance because of the rarity of these massive clusters. In order to understand which mass bins have more discriminating power, we define a normalized difference
\begin{equation}
	\epsilon=\frac{y_\mathrm{model}-y_\mathrm{fiducial}}{\sigma_y}, 
	\label{eq:signoise}
\end{equation}
where $\sigma_y$ is the dispersion in the observed HMF.  Lower mass bins have larger normalized difference $\epsilon$ because, though there is a large variance in the observed HMF, the nonfiducial model predictions shown lie outside of that variance.  However, the largest mass bin has smaller $\epsilon$, falling below $\epsilon=1$, meaning that the HMF with $\Mtot{}$ can no longer differentiate between the sample \model{} models at the 1-$\sigma$ level.

In addition to the true cluster masses, $\Mtot{}$, {for the Intermediate and Contaminated Catalogs, we calculate dynamical masses, $\Mdyn{}$, from galaxy velocity dispersion.  Because velocity dispersion as a mass proxy is both straightforward to apply and commonly used among observation \citep[e.g.][]{2013ApJ...772...25S, 2013JCAP...07..008H, 2014ApJ...792...45R, 2015ApJ...799..214B, 2016MNRAS.461..248S}, we use velocity dispersion as a mass proxy to obtain cluster dynamical masses.  The implementation of this} method is described in detail in \cite{Ntampaka2015} and \cite{Ntampaka2015b}, and is summarized {below} for reference.   {Clusters with dynamical mass $\Mtot{}\geq3.5\times10^{14}\,\Msolarh$ are used in the analysis of the Intermediate and Contaminated Catalog HMF.}

A traditional dynamical mass approach utilizes the virial theorem's prediction that cluster mass will scale with cluster member velocity dispersion, $\sigma_v$, as a power law.  The scaling relation can be written as
\begin{equation}
\sigma_{v}(\Mdyn{}) = \sigma_{15}\left( \frac{ \Mdyn{}}{10^{15} \, \mathrm{M_\odot}  h^{-1}} \right)^\alpha.
\label{eq:powerlaw}
\end{equation}
{with best fit power law parameters $\alpha$ and $\sigma_{15}$ (see Table \ref{table:catalog}) are} applied to predict mock cluster masses from LOS velocities.  The resulting mass predictions have approximately lognormal scatter about the true mass, with width $0.11\,\mathrm{dex}$.  {For the Intermediate and Contaminated Catalog HMF, g}alaxy velocities are given a conservative gaussian error of $100 \, \kms$ before dynamical masses are predicted, and {clusters with dynamical masses above $3.5\times10^{14}\,\Msolarh$ are used in the HMF}.  The scaling that is determined by the fiducial model and Multidark HMF is also applied to the $\Mdyn{}$ HMF.

Cluster mass estimates such as $\Mdyn$ have measurement errors, and introducing errors changes the shape and amplitude of the HMF.   Figure \ref{fig:hmf} shows that the resulting mass function, when uncorrected for the large scatter inherent in such a scaling relation, is inconsistent with the fiducial case.  The steeply-declining true mass function coupled with the large scatter in the $M$-$\sigma_v$ relationship yields an upscatter at high masses, creating a mass function that more closely resembles a high-\sig{}, low-\OM{} prediction.  

{Here, we have not attempted to correct for Eddington bias due to scatter in dynamical masses, though this would be a necessary step for performing a proper cosmological analysis using the HMF. Under the assumption of lognormal scatter, analytic approaches \citep[e.g.][]{2014MNRAS.441.3562E, 2011PhRvD..83b3015M} can correct for Eddington bias in the halo mass function.  Though lognormal scatter in dynamical masses is found when pure and complete cluster membership information is available \citep[e.g.][]{Evrard:2008aa}, interlopers contaminating the cluster observations produce scatter that is a function of true cluster mass \citep[e.g.][]{2013ApJ...772...47S, Ntampaka2015b}.  In this case, correcting for Eddington bias is nontrivial.  We will next present a method for collecting LOS velocities and directly using the resulting velocity distribution function as a cosmological probe that, unlike the HMF, is robust to the introduction of measurement errors.}   

\subsection{Velocity Distribution Function}

\begin{figure}[!t]
\begin{center}
\begin{tabular}{c c}
	\includegraphics[width=0.45\textwidth]{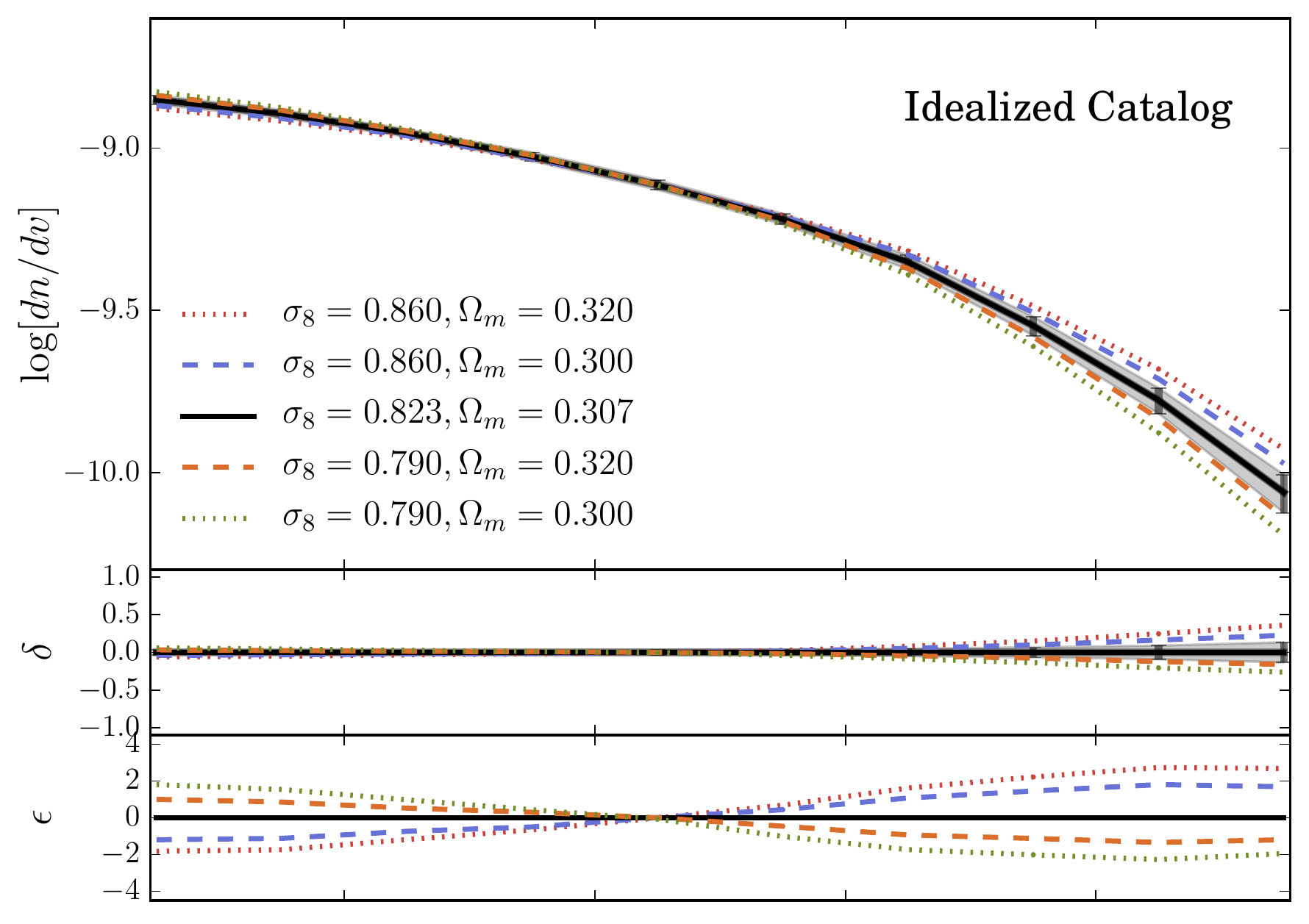} & \\
	\includegraphics[width=0.45\textwidth]{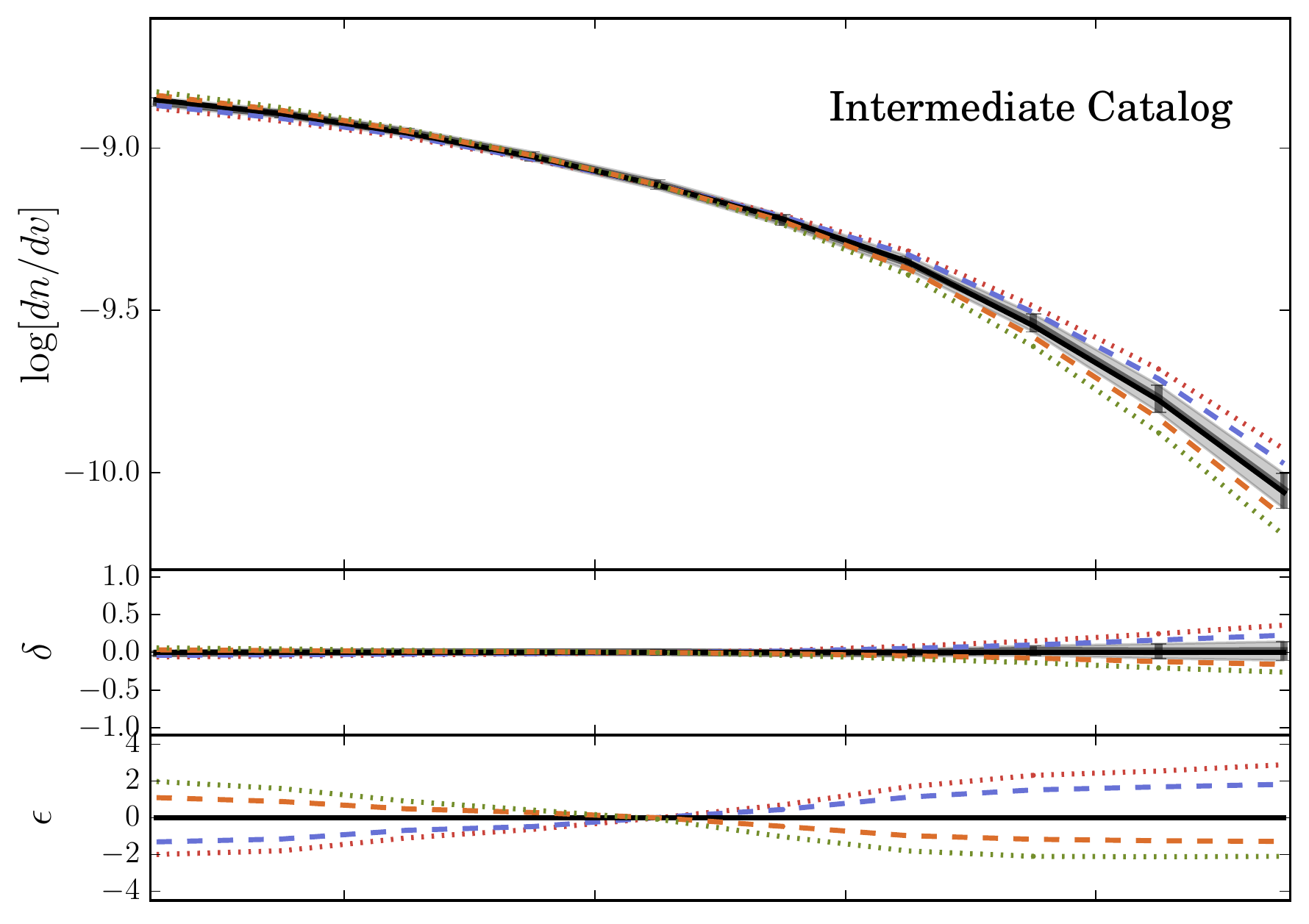} & \\
	\includegraphics[width=0.45\textwidth]{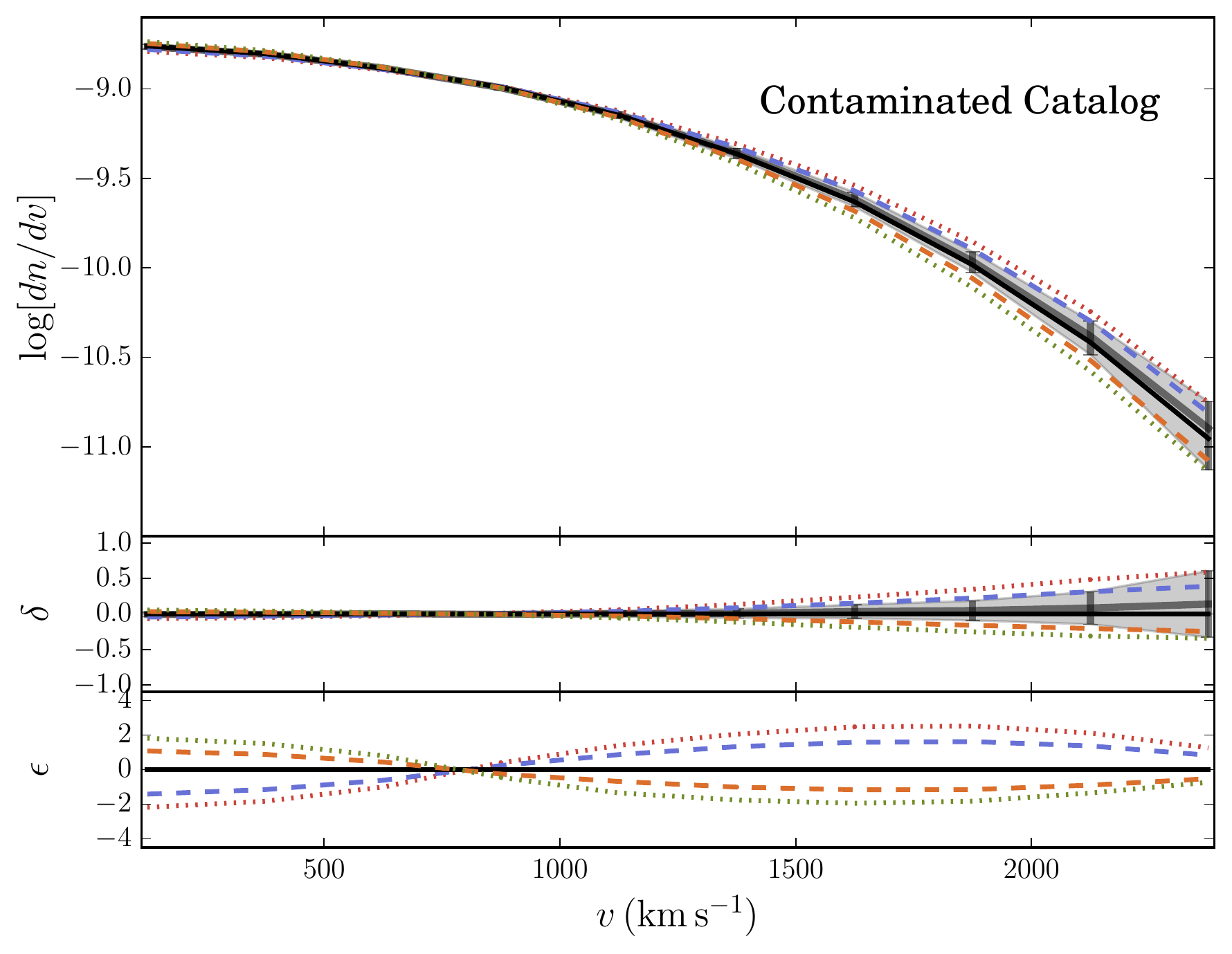} & \\
\end{tabular}
\end{center}
	\caption{\footnotesize{{Top:  VDF of the \new{Idealized} Catalog. For the fiducial model (black solid), the expected 68\% scatter (light gray) is shown for mock observations with 200 clusters. Four additional \model{} model predictions (colored dashed and dotted) are shown for comparison (top).  The fractional difference between models, $\delta$, is the largest at highest velocities (middle), while the normalized difference, $\epsilon$, peaks at \smash{$\approx 2000 \, \kms$}, where velocity bins have the most resolving power (bottom).
	Middle:  VDF of the Intermediate Catalog.  A conservative \smash{$100\,\kms$} Gaussian velocity error makes no significant effect on the VDF, demonstrating the statistical power of this new approach under the introduction of errors. 
	Bottom:  VDF of the Contaminated Catalog.  The velocity error introduces a slight offset between the fiducial model and the VDF at high velocities (dark gray) but the VDF is still well within 1-$\sigma$ of the fiducial model. \\ \\ } }}

	\label{fig:vdf}
\end{figure}

We define a velocity distribution function (VDF), a new statistic that can robustly be predicted with simulations which capture the dynamics of subhalos or galaxies.  Because the VDF is well-modeled by simulation, it can be used directly to explore constraints on cosmological parameters, circumventing the measurement errors associated with dynamical masses.  The VDF, $dn(v)/dv$, is a sum of PDFs of cluster member LOS velocities.  It is defined as
\begin{equation}
	\frac{dn}{dv} (v) \equiv \frac{1}{V}\sum_{i=1}^{N} \left[ \mathrm{PDF}(|v|) \right]_i,
	\label{eq:VDF}
\end{equation}
where $V$ is the volume of the region under consideration, $\mathrm{PDF}(|v|)$ denotes a probability distribution function of the absolute value of galaxy LOS velocities, and {the index $i$ denotes a sum over $N$ clusters.  These clusters may either be the most massive, denoted $dn(M,v)/dv$, or the richest, denoted $dn({\cal R},v)/dv$.  

{The halos in our sample are defined by an overdensity of $200\rho_\mathrm{crit}$, where $\rho_\mathrm{crit}$ is the critical density of the Universe.  To calculate $dn(v)/dv$ for the nonfiducial \sig{} and \OM{} cases, we must consider both the number of clusters of mass $M$ as well as their velocity scaling.  From the virial theorem, velocities should scale as a function of cluster mass $M$ and radius $R$, as
\begin{equation}
v \propto \left( \frac{M}{R}\right)^{1/2} \propto M^{1/3} \rho_\mathrm{crit}^{1/6}(z).
\label{vscale}
\end{equation}
For flat cosmologies at $z=0$, there is no normalization correction for velocities as a function of cluster mass.  Thus, equation \ref{eq:VDF} is weighted according to the number of clusters of mass $M_i$ predicted by a given cosmology with parameters \sig{} and \OM{},}
\begin{equation}
	\frac{dn}{dv} (v) \equiv \frac{1}{V}\sum_{i=1}^{N} \left[ w_i(M_i, \sigma_8, \Omega_m)\,  \mathrm{PDF}(|v|) \right]_i,
\end{equation}
where $w_i(M_i, \sigma_8, \Omega_m)$ is the ratio of the nonfiducial analytic HMF, $dn/d\log(M)$, to the fiducial HMF evaluated at $M_i$, the $i^{th}$ cluster mass.  The number of clusters contributing in the nonfiducial case is chosen so that $\sum w_i=N$.

We choose $0\,\kms$ as the lower velocity limit and as with the HMF, choose an upper velocity limit where the counts to bin dispersion goes below 3.  As with the HMF, the VDF bin sizes are chosen according to Freedman-Diaconis' rule (Equation \ref{eq:FD}) {applied to the Contaminated Catalog velocities, with $\{x\}=\{v\}$ and $N=200$. This yields 10 bins from $|v|=0\,\kms$ to $|v|=2500\,\kms$ with $dv=250\,\kms$.}

Figure \ref{fig:vdf} shows the Multidark VDF.  The true VDF is shown in comparison to a representative sample of other \model{} VDFs.  Because changing \sig{} or \OM{} changes the number of massive halos and the shape of the HMF, it similarly changes the shape of the VDF.  Higher-mass halos have larger velocity dispersions, so \model{} models that predict a large number of massive clusters predict a large number of high velocity galaxies as well. The resulting VDF is shallower near $v=0$ and has an increased number density at higher velocities.

The anticorrelation between high and low velocities is most evident in normalized difference $\epsilon$, where the five different VDFs all cross the $\epsilon=0$ line at $v\approx1200\,\kms$.  This velocity corresponds to the velocity dispersion of a typical $10^{15}\,\Msolarh$ cluster, and clusters with smaller masses will tend to contribute more in the $|v|<1200\,\kms$ region of the VDF.  

To mimic what might happen when the VDF is applied to observation, velocities with error, $\verr{}$, are also considered.  As a counterpart to the HMF with $\Mdyn{}$
in which cluster masses are given a realistic error, here every measured velocity is given an error.  Galaxy LOS velocities are given scatter according to a Gaussian distribution with a conservative error estimate, $\sigma_\mathrm{gauss}=100 \kms$.  

Figure \ref{fig:vdf} shows that this introduction of {velocity} error has little effect on the VDF, with deviations from the fiducial model only becoming evident at the highest velocities where there are few galaxies.  {Introducing interlopers, however, alters the shape of the VDF, with many fewer high-velocity members.  In practice, these interloping galaxies would be modeled into the VDF.}  The VDF assembled from the 200 most massive clusters is shown; the VDF with the 200 richest clusters has a qualitatively similar shape, amplitude, $\delta(v)$, and $\epsilon(v)$.  The VDF with fewer clusters is similar in shape and amplitude, though variance is larger due to the smaller mock observation.  Unlike in the HMF with $\Mdyn{}$, the introduction of reasonable errors to the VDF is still within 1-$\sigma$ agreement of the fiducial case across bins.  

{We note that the use of cluster member LOS velocities as a cosmological probe is not new.  Velocity PDFs have been used to probe the distribution of dark matter in clusters \citep{1987ApJ...313..121M} and velocity dispersion counts have been used to constrain cosmological parameters \citep{Evrard:2008aa, 2016arXiv160200611C}{.  Additionally, the distribution of velocity dispersions has been used to probe the bias between the velocity dispersions of observed cluster galaxies and simulated cluster dark matter \citep{1996A&A...310...31M}. }  New to our method is the quantifying of galaxy clusters with the distribution of LOS velocities, and using this distribution as a direct probe of cosmological models.}

\section{{Results}}
\label{sec:results}

 \begin{figure*}[!th]
\begin{center}
\begin{tabular}{c c}
        	\includegraphics[width=0.5\textwidth]{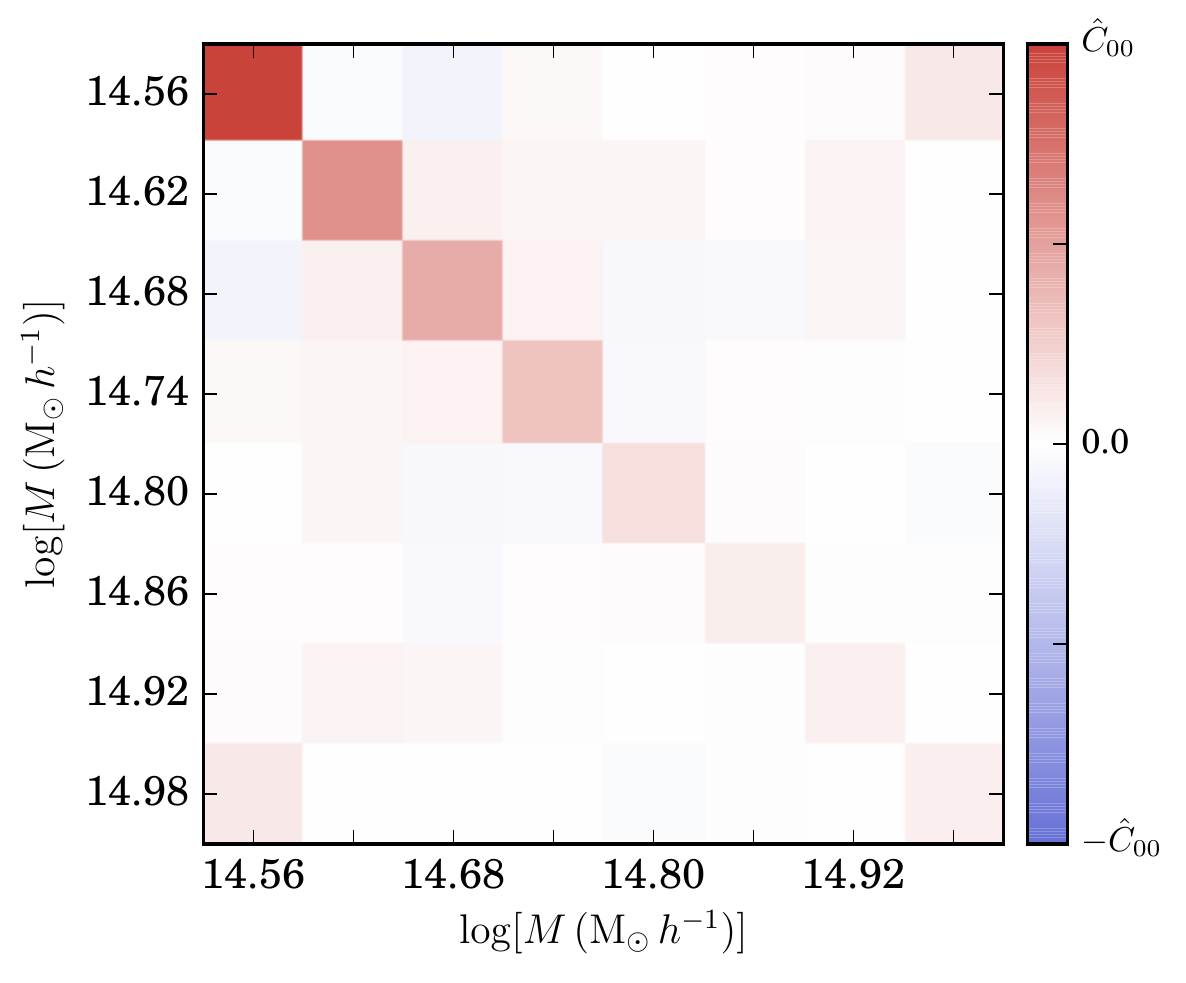} & \includegraphics[width=0.5\textwidth]{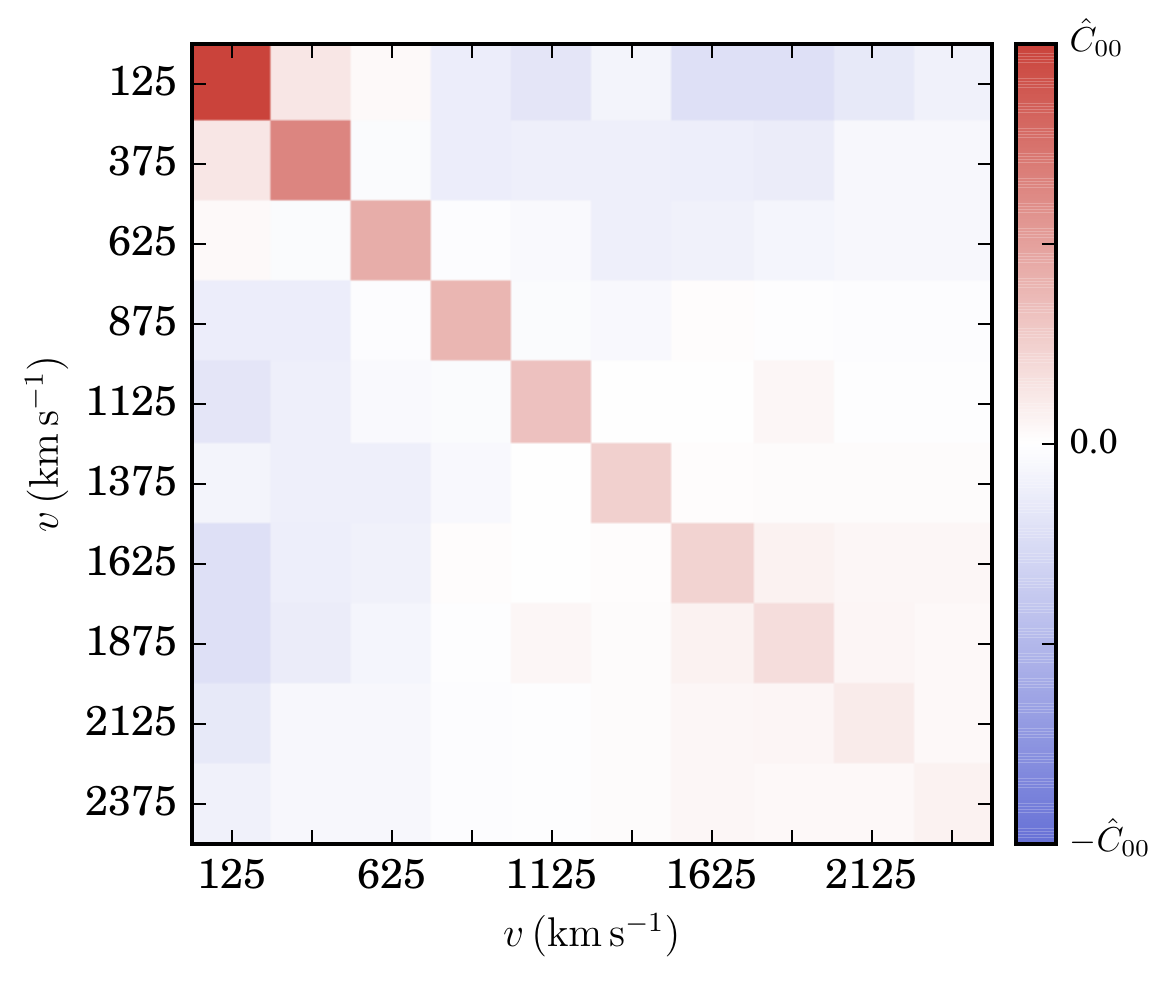} \\
	\includegraphics[width=0.5\textwidth]{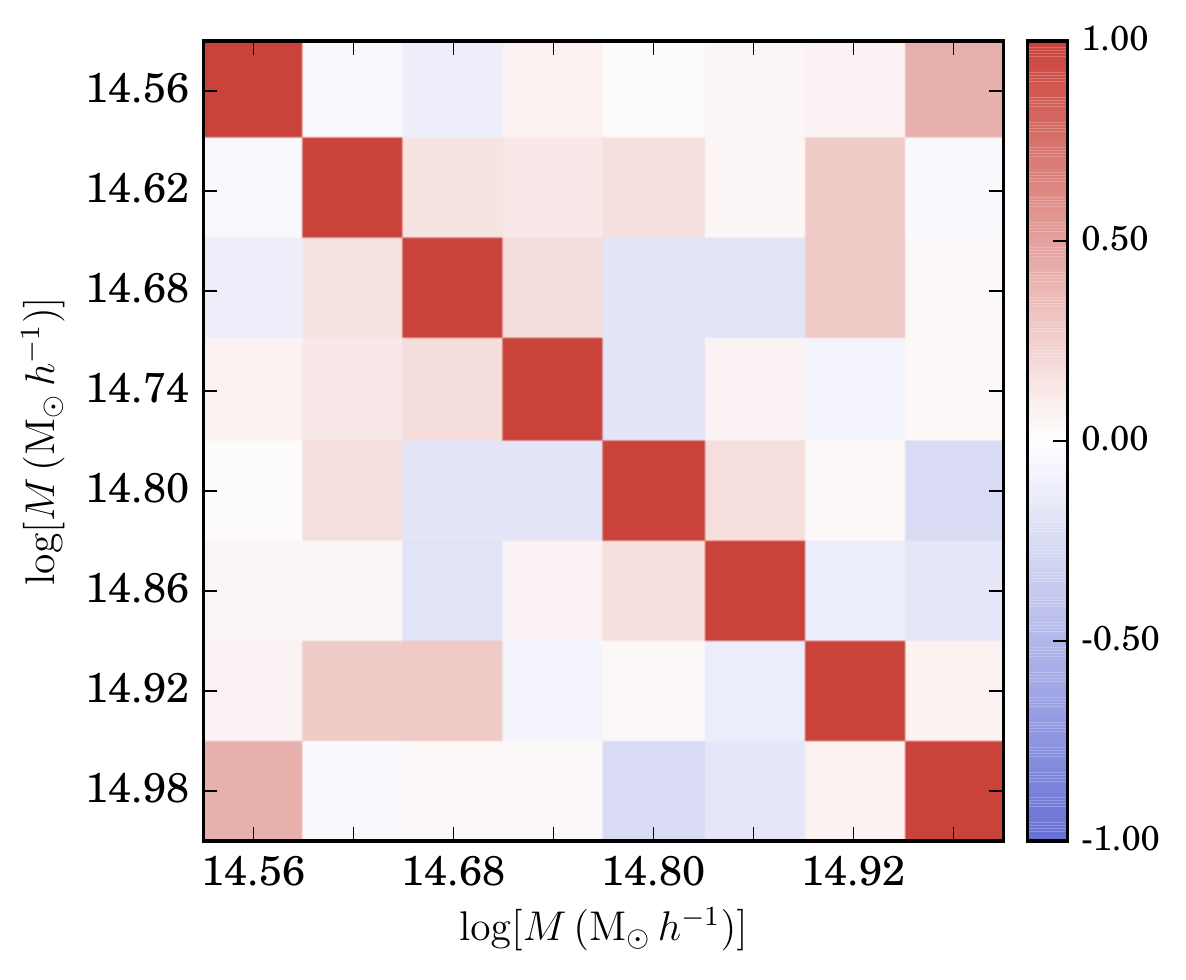} & \includegraphics[width=0.5\textwidth]{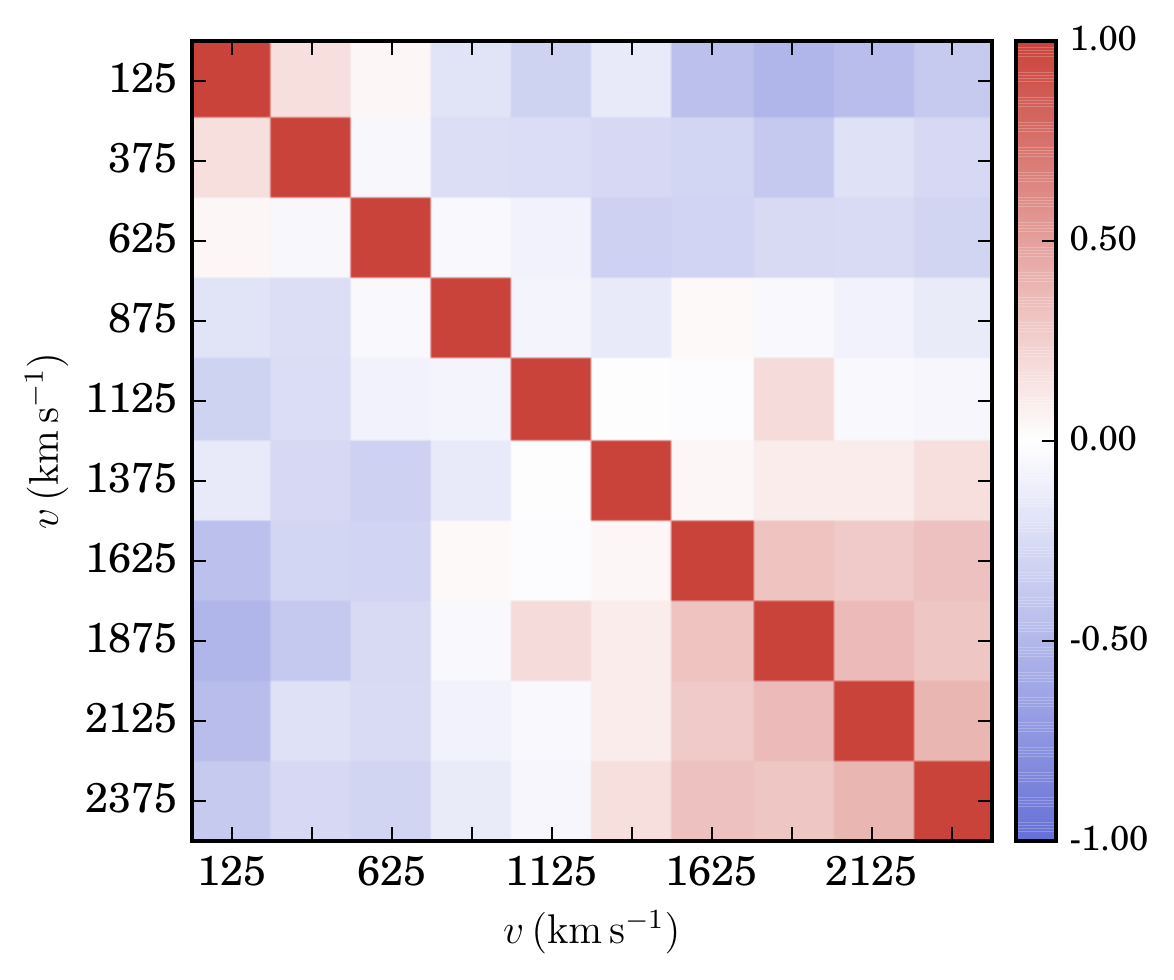} \\
\end{tabular}
      	  \caption{Left: {\new{Idealized} Catalog} HMF covariance matrix (\smash{$\hat{C}$}, top) and correlation matrix ($C_{i,j}/(\sigma_i \sigma_j)$, bottom).  Low-mass halos are most common and also have greatest variance in their counts.  Halo counts are positively correlated across mass due to large-scale density fluctuations.  Right: {\new{Idealized} Catalog} VDF covariance matrix (top) and correlation matrix (bottom).  {The bins above $\approx1125\,\kms$ are} heavily influenced by the number of high-mass clusters in any given observation.  The positive cross-correlation among the highest-velocity bins are due to a small number of high-mass clusters contributing these velocities.  Similarly, the smallest-velocity bins are primarily constructed from abundant low-mass clusters.   The negative covariances are due to the counting nature of the VDF; observations with an abundance of high-mass clusters will necessarily have fewer low-mass clusters contributing to the VDF.  \\
	 }

       	\label{fig:covariance}
      	\end{center}
\end{figure*}

 \begin{figure*}[!th]
\begin{center}
\begin{tabular}{c c c}
        	\includegraphics[width=0.33\textwidth]{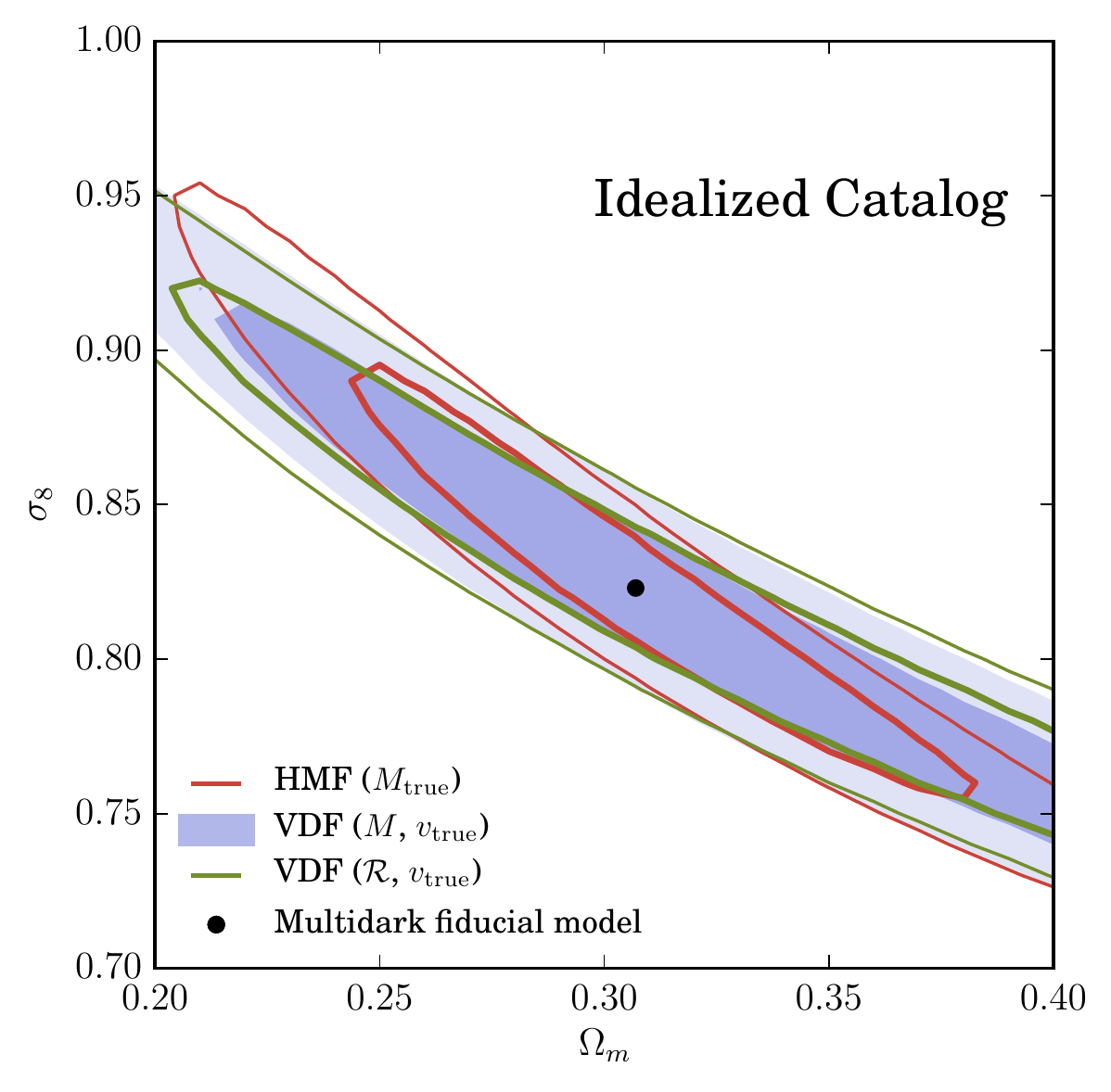} &\includegraphics[width=0.33\textwidth]{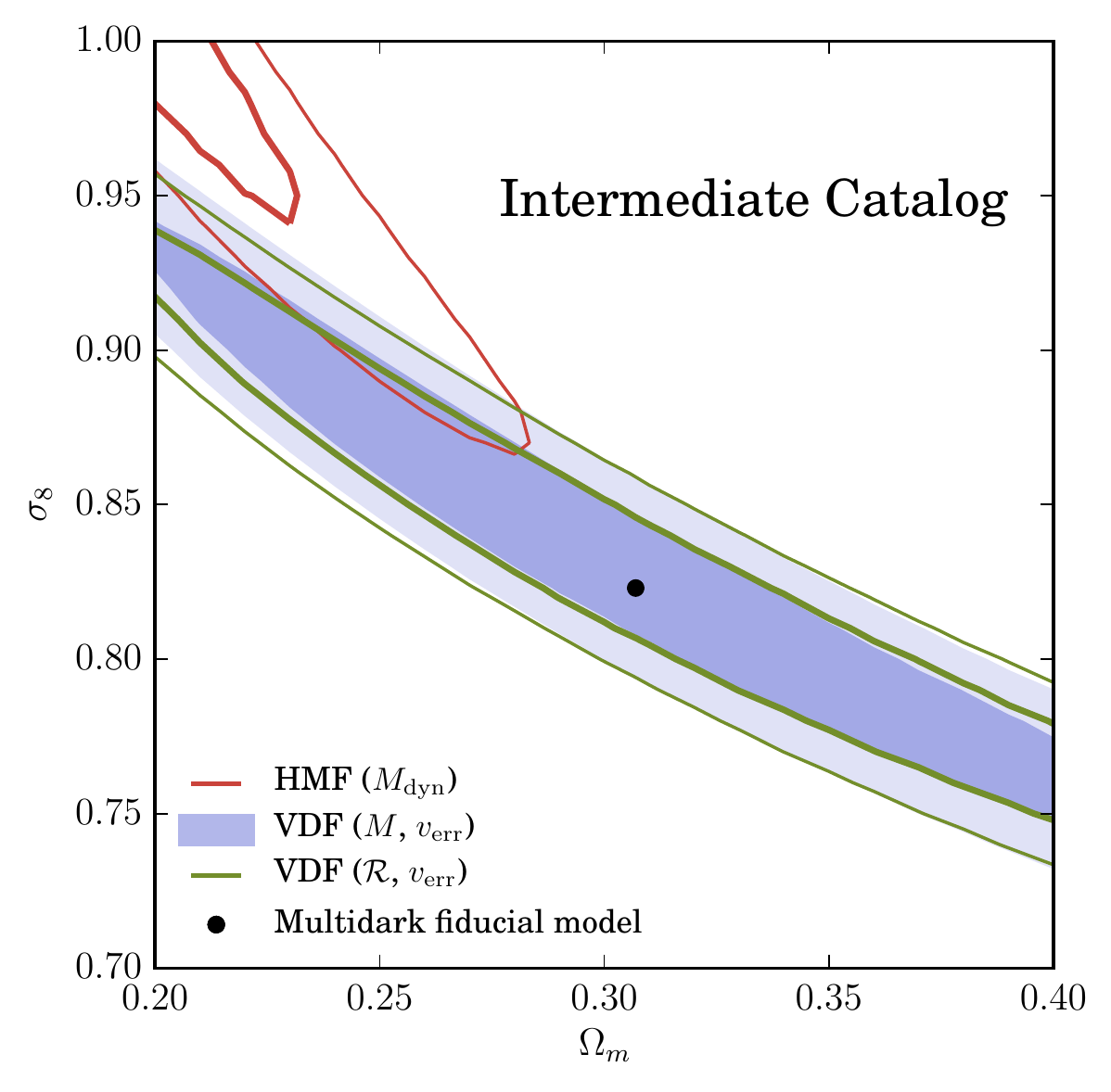}  &\includegraphics[width=0.33\textwidth]{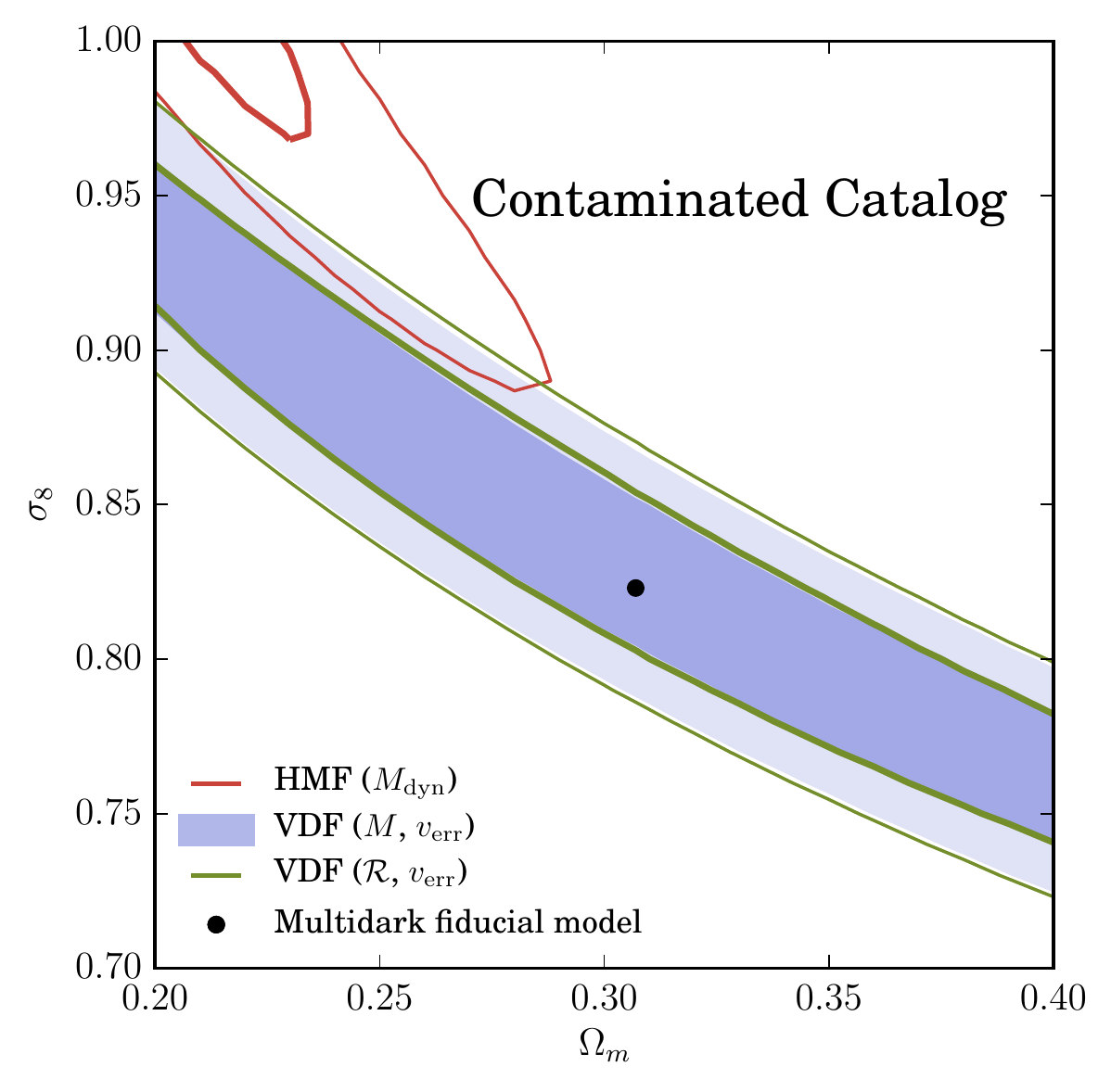}\\
	\includegraphics[width=0.33\textwidth]{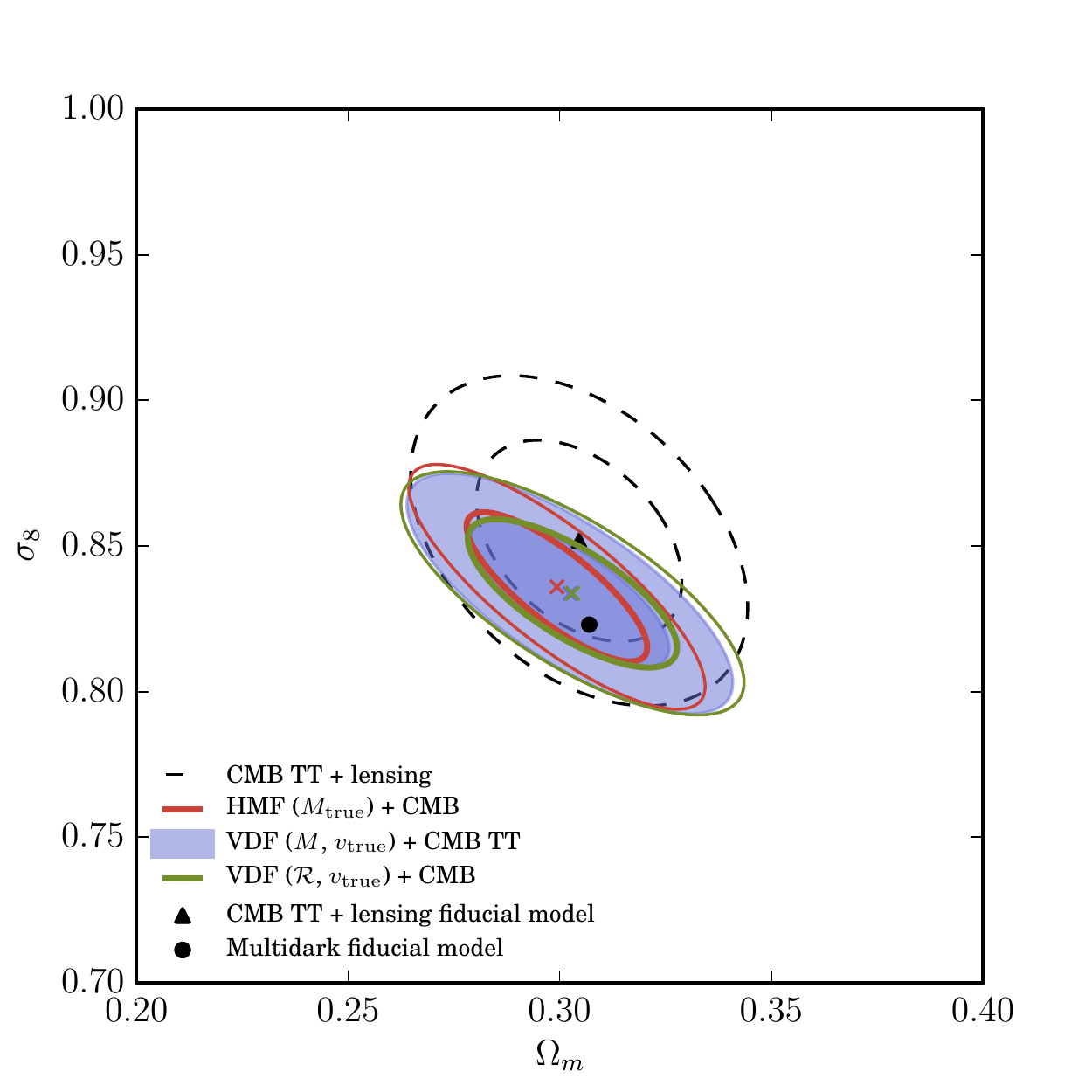} &\includegraphics[width=0.33\textwidth]{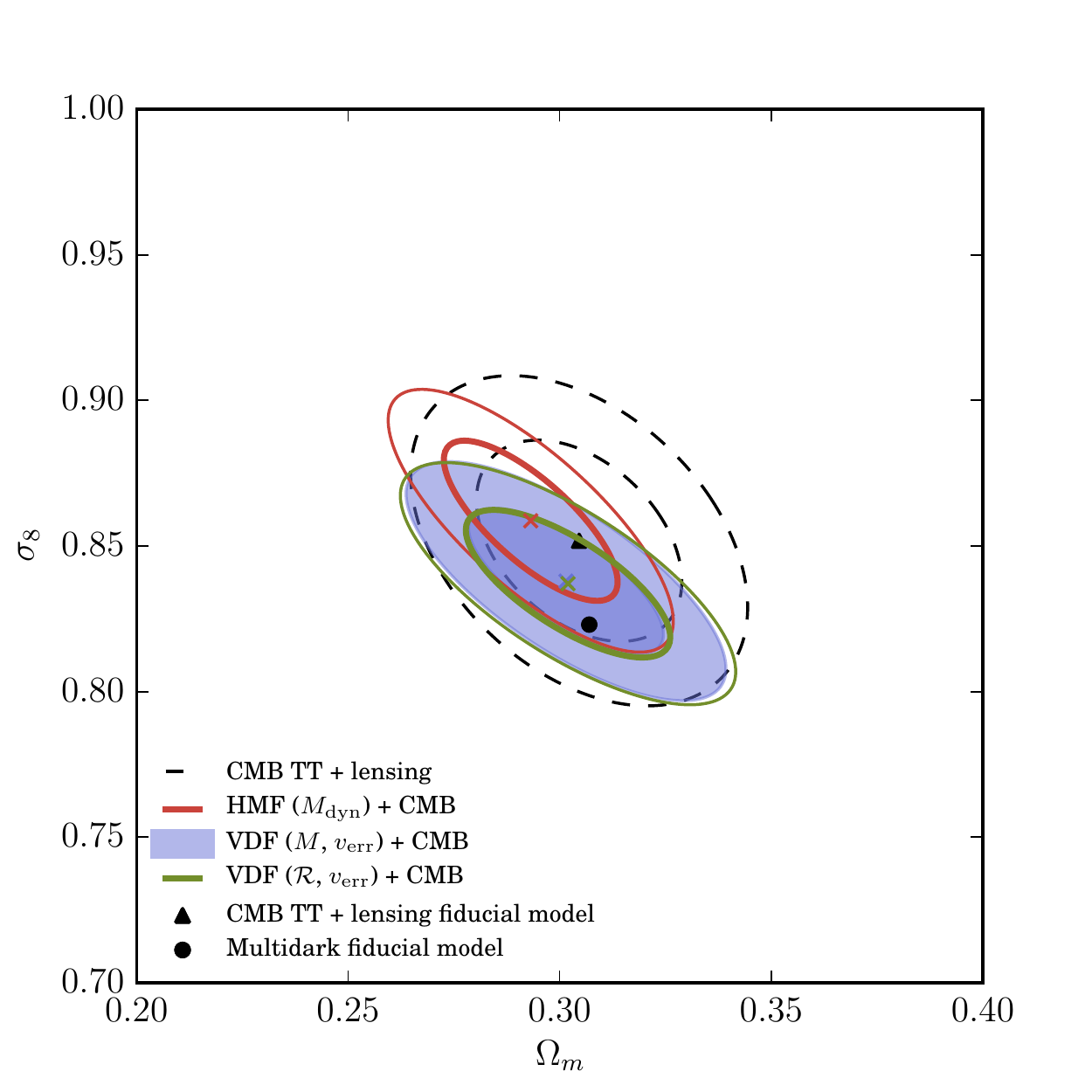} &\includegraphics[width=0.33\textwidth]{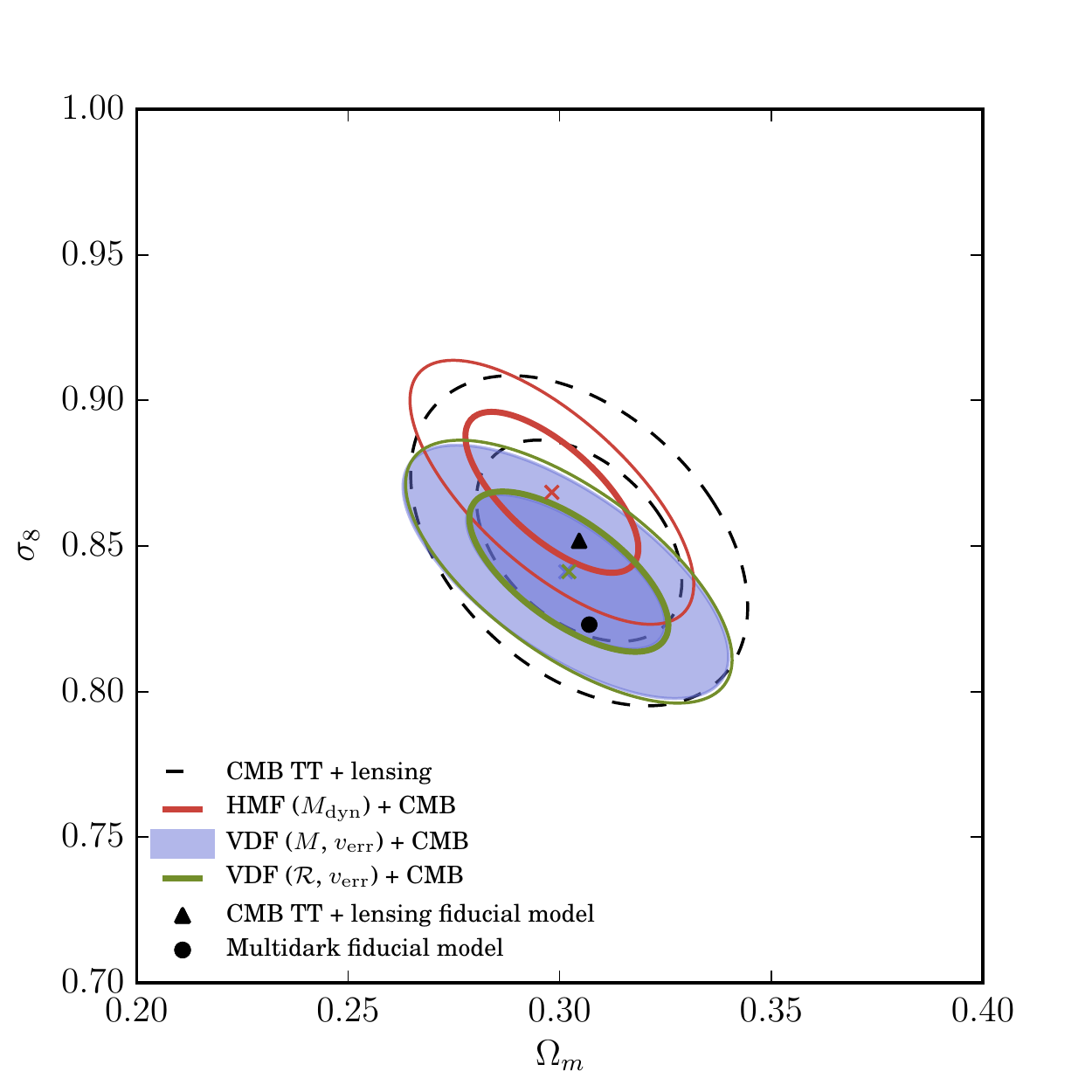} \\
\end{tabular}
      	 \caption{Left:  68- and 95\% confidence regions (contours) of the HMF (red), VDF with mass sorting (blue), and VDF with richness sorting (green) {for the \new{Idealized} Catalog.}  When combined with CMB TT constraints (black dashed), these techniques produce similar maximum likelihoods (colored x scatter points) and constraints on $\sigma_8$ and $\Omega_m$ (bottom). 
	 {Middle}:  Confidence regions and maximum likelihoods \new {for the Intermediate Catalog, which includes velocity errors and dynamical mass estimates}.  The HMF with dynamical masses (red), if uncorrected for scatter and bias, produces confidence regions centered on low $\Omega_m$ and high $\sigma_8$, in more than 2-$\sigma$ disagreement with the Multidark fiducial model.
	 {Right:  The inclusion of interlopers in the Contaminated Catalog causes the HMF to prefer a high \sig, low \OM{} cosmology.  The VDF constraints widen compared to a more ideal case, but are minimally biased.}
	 }
       	\label{fig:likelihood}
      	\end{center}
\end{figure*}

\subsection{Covariance Matrices}

The HMF and VDF of mock observations can be used to forecast constraints on \sig{} and \OM{}.  This is accomplished by comparing the observed HMF or VDF to non-fiducial predictions by a standard $\chi^2$ analysis.  {The \new{Idealized}, Intermediate, and Contaminated Catalog mock HMFs are compared to the analytic HMF given by \cite{2008ApJ...688..709T}.  The \new{Idealized} and Intermediate VDFs are compared to the mean \new{Idealized} VDF.  The Contaminated VDF is compared to the mean VDF constructed with the Contaminated Catalog, but without velocity errors.}  The estimated covariance matrix, \smash{$\hat{C}$}, is given by
\begin{equation}
\hat{C}= \frac{1}{n_\mathrm{obs}-1} \sum_{i=1}^{n_\mathrm{obs}} \left[ (y_i-\bar{y})(y_i-\bar{y})^T\right],
\label{eq:cov}
\end{equation}
where $y_i$ is a $n_\mathrm{bin}\times1$ column vector of the $i^{th}$ observation's $n_\mathrm{bin}$ bin values, $\bar{y}$ is a $n_\mathrm{bin}\times1$ column vector of the mean observation, $n_\mathrm{obs}$ is the number of independent observations, and $n_\mathrm{bin}$ is the number of bins.

The resulting covariance matrices are shown in Figure \ref{fig:covariance}.  Also shown are the closely-related correlation matrices, where the ($i$, $j$) term of \smash{$\hat{C}$} is normalized by $\sigma_i \sigma_j$.  Because the covariance matrix spans a wide range of values, such a normalization can highlight smaller covariances and provide an intuition for the nature of the covariance matrix.  The smallest bin in the HMF covariance matrix has the largest variance, as is evident in Figure \ref{fig:hmf}.  Some positive correlation is evident between mass bins, a remnant of large-scale density fluctuations which resulted in a larger average density in some mock observations and lower average density in others, compared to the average cosmic density.

The off-axis positive and negative covariances of the VDF \smash{$\hat{C}$}, however, have a different origin.  While positive HMF covariance originated from large-scale density fluctuations, the positive covariances in the VDF are an effect of the counting.  Massive halos generally have a larger velocity dispersion, and mock observations with an overabundance of massive halos will tend to have an overabundance of high-velocity galaxies, leading to the positive correlations among velocity bins.  Due to hard limit on the number of clusters used to calculate the VDF, however, an abundance of massive halos necessarily leads to a deficit of low-mass halos included in the $N$ most massive or richest clusters that contribute to the VDF.  This is the cause of the regions of negatively-correlated velocities.

As detailed in, e.g., \cite{2007A&A...464..399H, 2013MNRAS.432.1928T, 2014MNRAS.439.2531P}, an unbiased estimator, $\hat{\Psi}^{-1}$, of the inverse covariance matrix is given by
\begin{equation}
\hat{\Psi}^{-1} = \frac{n_\mathrm{obs}-n_\mathrm{bin}-2}{n_\mathrm{obs}-1} \hat{C}^{-1},
\label{eq:invcov}
\end{equation}
where $\hat{C}^{-1}$ denotes a standard matrix inversion of the covariance matrix.  The number of observations sets a maximum for the number of bins that can be considered; the number of bins is constrained to $n_\mathrm{bin}< n_\mathrm{obs}-2$. \\ \\

\subsection{Constraints on \sig{} and \OM{}}

  The posterior probability, $P(\mathmodel{}|y)$, of a model given the HMF or VDF, generically denoted $y=\{\bar{y},\Psi^{-1}\}$, can be determined by 
\begin{equation}
P(\mathmodel{}|y)\propto\mathcal{L}(y|\mathmodel{}),
\label{eq:posterior}
\end{equation}
where the likelihood is
\begin{equation}
\mathcal{L}(y|\mathmodel{}) \propto \exp{\left(\frac{-\chi^2(y|\mathmodel{})}{2}\right)},
\label{eq:likelihood}
\end{equation}
and 
\begin{equation}
\chi^2(y|\mathmodel{}) = (\bar{y}-y^\star)^T \, \hat{\Psi}^{-1} \, (\bar{y}-y^\star),
\label{eq:chi2}
\end{equation}
under the assumption of uniform prior probability for \sig{} and \OM{}.  Here, $y^\star$ is a $n_\mathrm{bin}\times1$ column vector of the \model{} model's predictions for the bin center.

Figure \ref{fig:likelihood} shows the 68- and 95\% confidence regions in the \model{} plane.  The confidence regions are determined by a smooth curve that lies where $\Delta \chi^2$ from the maximum $\mathcal{L}$ (minimum $\chi^2$) is at the expected value for a $\chi^2$ distribution, given by
\begin{equation}
	\Delta\chi^2  = \chi^2_{1\sigma}-\chi^2_\mathrm{min} = F^{-1}_{2}(0.6827),
	\label{eq:deltaChi2}
\end{equation}
where $F^{-1}_{2}(0.6827)$ is the inverse cumulative distribution function of a $\chi^2$ distribution with two degrees of freedom, one each for \sig{} and \OM{}, evaluated at 68.27\%.  A similar curve for the 95.45\% confidence region is set by \smash{$\Delta\chi^2=F^{-1}_{2}(0.9545)$}.  The error-free VDF with $\vtrue{}$ produces comparable confidence regions near the fiducial model to its mass function counterpart, the HMF with $\Mtot$.  However, the introduction of errors to velocity combined with mass estimate $\Mdyn$ causes a significant offset in the HMF, while the 
VDF remains largely unchanged.

{Table \ref{table:summary} lists the smallest contour, $c$, containing the Multidark fiducial model ($\sigma_{8, \, \mathrm{fiducial}}=0.823$ and $\Omega_{m, \, \mathrm{fiducial}}=0.307$). Inverting Equation \ref{eq:deltaChi2} and evaluating it at the fiducial model gives this contour,
\begin{equation}
c = F_2(\chi^2_\mathrm{fiducial}-\chi^2_\mathrm{max\,\mathcal{L}}),
\label{eq:contour}
\end{equation}
where $\chi^2_\mathrm{fiducial}$ is the $\chi^2$ value at the Multidark fiducial model, $\chi^2_\mathrm{max\,\mathcal{L}}$ is the maximum likelihood $\chi^2$, and $F_2$ is the cumulative distribution function of a 2-dimensional $\chi^2$ distribution.  Increasing the volume of the observation tightens the constraints in the HMF and VDF cases, and the fiducial model therefore lies on a larger contour.  The bias, however, is the dominant contributor to the difference between the VDF and HMF $c$ values; while the fiducial model is in good agreement with the VDF, the HMF is biased to low \OM{} and high \sig{} and the fiducial model lies well outside of the 68\% contour for all observation sizes.}

The VDF and HMF 68\% constraints can be approximated as a band in the \model{} plane, parameterized as
\begin{equation}
\sigma_8 \, \Omega_m^\gamma = \mathcal{A}\pm\delta\mathcal{A}_{68\%},
\label{eq:band}
\end{equation} 
where $\sigma_8 \, \Omega_m^\gamma = \mathcal{A}$ describes a maximum likelihood ridge in the \model{} plane and $\delta\mathcal{A}_{68\%}$ gives the distance from that ridge to the contours containing the 68\% likelihood region.  The 95\% likelihood region can be similarly described, though it should be noted that  $\delta\mathcal{A}_{95\%}\neq2\times\delta\mathcal{A}_{68\%}$.

At all $N$, the width of the VDF 68\% confidence region, $\delta\mathcal{A}_{68\%}$, has a negligible increase when reasonable errors are added to LOS velocities.  For example, with $N=200$ clusters, VDF($M$, \new{Idealized}) constrains the parameter combination \updated{\smash{$\sigma_8\,\Omega_m^{0.30} =0.579 \pm 0.011 $}}.  Adding velocity errors, VDF($M$, Intermediate), provides similar constraints with only minor bias:  \updated{\smash{$\sigma_8\,\Omega_m^{0.29} = 0.584 \pm 0.011 $}}. {Even with the inclusion of interlopers, VDF($M$, Contaminated), the constraints widen only slightly  \updated{\smash{$\sigma_8\,\Omega_m^{0.29} = 0.589 \pm 0.014$}.}   Likewise, adding velocity errors and richness sorting, VDF($\mathcal{R}$, Intermediate),  similarly constrains in the \model{} plane: \updated{\smash{$\sigma_8\,\Omega_m^{0.28} = 0.594 \pm 0.011$}}.   }  

In contrast, adding {velocity errors and interlopers} changes maximum likelihood ridge parameters $\gamma$ and $\mathcal{A}$ significantly in addition to increasing the width of the 68\% confidence band, $\delta\mathcal{A}_{68\%}$.  With $N=200$, the {\new{Idealized} Catalog} HMF fit changes from \updated{\smash{$\sigma_8\,\Omega_m^{0.37} = 0.531 \pm 0.009$}} to \updated{\smash{$\sigma_8\,\Omega_m^{0.42} = 0.510 \pm 0.009$}} {for the Intermediate Catalog.}  {The Contaminated Catalog constraints are biased further, to \updated{\smash{$\sigma_8\,\Omega_m^{0.44} = 0.515 \pm 0.011 $}} from the inclusion of interlopers.}

As can be seen in Table \ref{table:summary}, VDF $\delta\mathcal{A}_{68\%}$ decreases with sample size, as \updated{$\approx N^{-0.3\, \mathrm{to} \, -0.6}$}.  This decreasing width of the 68\% constraints leads to tighter constraints in the \model{} plane with increasing $N$.  The fiducial model lies well within the 68\% constraints of the VDF with $N=50$, $100$, and $200$.  While the HMF $\delta\mathcal{A}_{68\%}$ similarly decreases with sample size, the bias with $N=200$ clusters results in the fiducial model being excluded by HMF constraints at more than \updated{99\%} { for the Intermediate and Contaminated Catalogs}.

\subsection{Joint Constraints with CMB}

The band-shaped HMF and VDF likelihoods are combined with constraints obtained from the Planck CMB temperature-temperature (TT) correlation data for multipoles $\ell\leq 2508$, \citep{2015arXiv150201589P}, significantly reducing the degeneracy between \sig{} and \OM{}.  The joint posterior probability for \sig{} and \OM{} is related to the Planck posterior \citep{2014A&A...571A..16P} by
\begin{align}
\begin{split}
	P(\sigma_8, \Omega_m|y, \mathrm{CMB}) &= \frac{P(\sigma_8, \Omega_m|y)\, P(\sigma_8, \Omega_m|\mathrm{CMB})}{P(\sigma_8, \Omega_m)}\\
	\\
	&=\mathcal{L}(y|\sigma_8,\Omega_m)  P(\sigma_8, \Omega_m|\mathrm{CMB}),\\
	\end{split}
	\label{eq:jointL}
\end{align}
where $P$ denotes a probability distribution function and $P(\sigma_8, \Omega_m)$ is a flat joint prior probability distribution on \sig{} and \OM{}.  The resulting probability $P(\sigma_8, \Omega_m|y, \mathrm{CMB})$ is assumed to be well-described by a bivariate Gaussian.

Because CMB TT constraints are centered on a different maximum likelihood than the Multidark simulation, offsets in joint constraints are introduced not only by velocity errors and mass estimates, but also by the two different fiducial models at play.  The joint constraint offset in maximum likelihood  \sig{} is compared before and after introducing velocity errors and mass estimates, as
\begin{equation}
\Delta\sigma_8=\sigma_{8, \, \mathrm{max}\mathcal{L}}-\sigma_{8, \, \mathrm{effective}},
\label{eq:DeltaCMB}
\end{equation}
where $\sigma_{8, \, \mathrm{max}\mathcal{L}}$ is the maximum likelihood after combining {the Intermediate or Contaminated HMF or VDF} with CMB constraints.  The effective fiducial model, $\sigma_{8, \, \mathrm{effective}}$ is the maximum likelihood $\sigma_8$ from {the \new{Idealized} catalog HMF or from the \new{Idealized} or Contaminated VDF}.  Thus, the {\new{Idealized} VDF} maximum likelihood serves as the effective \sig{} and \OM{} for {the Intermediate VDF}, {the VDF with interlopers but without velocity errors serves as the effective \sig{} and \OM{} for the Contaminated VDF, and the \new{Idealized} HMF} serves for {both the Intermediate and Contaminated} HMF.
An analogous calculation is performed for $\Delta\Omega_m$.  Again, we see that the HMF with dynamical mass estimates introduces the largest offset in \sig{} and \OM{}.  These are summarized in Table \ref{table:summary}.

Table \ref{table:summary} also summarizes the offsets and errors found when applying the VDF with velocity errors as well as the HMF with mass estimates.  The size of the 1-$\sigma$ error bars, ${(\sigma_8)}_{68\%}$ and ${(\Omega_m)}_{68\%}$, quantify the constraints on \sig{} and \OM{}, respectively.  While the CMB TT constraints alone give ${(\sigma_8)}_{68\%}=0.0228$ and ${(\Omega_m)}_{68\%}=0.0160$, the HMF and VDF provide comparable further constraints on these cosmological parameters, though the HMF maximum likelihood regions are biased to low \OM{} and high \sig{}.
The HMF and VDF of the smaller mock observations are quantitatively similar; the primary difference in these smaller observations occurs in the covariance matrix  \smash{$\hat{C}$}, which has larger variances due to smaller sample size.  As expected, bin variances (diagonal terms in \smash{$\hat{C}$}) scale as \smash{$\sqrt{N}$}, leading to tighter constraints with larger observations. Properly evaluating even larger mock observation and the resulting constraints would require a much larger simulation on which to evaluate \smash{$\hat{C}$}.

Both before and after combining with CMB constraints, adding reasonable {velocity errors and interloping galaxies causes a significant offset in the HMF} maximum likelihood from the fiducial value.  The VDF approach is less affected by the introduction of {velocity errors or interlopers}.  Choosing the richest clusters with VDF(${\cal R}$), compared to the most massive clusters with VDF($M$), has little effect on the resulting VDF constraints.

\section{Discussion}
\label{sec:discussion}

While cluster counts as a function of mass can be used to constrain cosmological parameters, a major source of error in the HMF approach is in connecting cluster observables with halo mass  and understanding bias and scatter in this relationship.  We have shown that collecting cluster member velocities in the form of the VDF is one viable alternative to constraining cosmological models with dynamical mass measurements.  While the VDF's constraining power is comparable to the HMF, its primary strength as a tool for evaluating cosmological models is its resiliency to the introduction of reasonable measurement errors.

Though Eddington bias contributes in both the HMF and VDF cases, it results in a large offset in the HMF constraints.  The large scatter in $M$ in comparison to bin width, the steeply-declining HMF, and the relatively few objects contributing to the HMF are three primary reasons for this.  {Correcting for this Eddington bias is nontrivial in the realistic case where interlopers contaminate the sample, creating a non-lognormal scatter that changes as a function of true halo mass.}  In our likelihood analysis, we have not attempted to account for Eddington bias in the analysis of the HMF, however, this step would be necessary to perform a proper cosmological analysis with dynamical masses.  The VDF, however, is relatively robust to scatter in the observable, $v$.  This scatter does not produce a large bias in \sig{} and \OM{}, even without correcting for scatter in the observable.   

Details about the observation to be analyzed and the expected sample selection, for example volume and mass or luminosity limits, should be taken into account in the modeling of the VDF.  Even velocity errors may be included in the modeling, though for velocity errors $\leq100\,\kms$, this may be unnecessary.  {We have shown that both mass and richness serve as adequate methods for choosing clusters to contribute to this statistic and the VDF constraining power is unaffected by which of these method is employed for choosing clusters.}  

However, we stress that the details of the observation must be captured by the modeling{, and that care should be taken to match survey selection effects.  For example,} VDF($M$) and VDF($\mathcal{R}$) take slightly different forms and cannot be substituted for one another.  Selecting clusters for VDF analysis may be achieved by applying
a mass proxy such as SZ, X-ray, or weak lensing masses.  While selecting clusters by velocity dispersion may introduce a bias in the VDF, the machine learning technique from \cite{Ntampaka2015} and \cite{Ntampaka2015b} is a viable alternative for selecting clusters to contribute to the VDF from galaxy LOS velocities; this method reproduces the true HMF more accurately than a dynamical mass approach.

Applying the VDF as a cosmological probe requires a large spectroscopic data set of cluster members.  Such data exist, and future surveys will probe larger volumes and fainter magnitudes.   The Hectospec Cluster Survey (HeCS) has spectra of 21,314 galaxies, 10,275 of which are galaxy cluster members \citep{2013ApJ...767...15R}.     Upcoming surveys include the Hobby-Eberly Telescope Dark Energy Experiment (HETDEX), which expects to observe 0.8 million Lyman-$\alpha$ galaxies in the next three years \citep{2011ApJS..192....5A, 2014SPIE.9147E..0QH},  
as well as the Subaru telescope Prime Focus Spectrograph (PFS), having scheduled first light in 2018 \citep{2015arXiv150700725S}.

Other spectroscopic surveys with smaller sample size or lower velocity resolutions may also be utilized for preliminary tests of the VDF method.  Spectroscopic follow up to clusters detected by the Atacama Cosmology Telescope \citep{2015arXiv151200910S} and South Pole Telescope \citep{2014ApJ...792...45R} currently have data for 44 and 48 clusters, respectively.  Advanced ACTPOL \citep{2015arXiv151002809H} and SPT-3G \citep{2014SPIE.9153E..1PB} will detect more clusters for spectroscopic followup.

Space-based telescope Euclid will measure redshifts of 50 million galaxies out to $z>2$ \citep{2011arXiv1110.3193L}, and the Dark Energy Spectroscopic Instrument (DESI) will have spectra for 18 million emission-line galaxies during its 2018-2022 science run \citep{2013arXiv1308.0847L}, though both have velocity errors larger than the error explored in this work.  Exploring the effect of larger velocity errors is a necessary step before applying the VDF technique to these observations.

Current and future observations will produce large catalogs of galaxy velocities.  While the adopted volume used in the present analysis is limited by the simulation volume, changing sample size follows the expected trend: larger observations lead to decreased bin dispersion, scaling as \smash{$\sqrt{N}$}, which in turn lead to tighter constraints.  Though smaller cluster surveys may not provide \sig{} and \OM{} constraints competitive with CMB TT + polarization constraints, large volume surveys can be utilized in conjunction with the VDF to test cosmological models.

While a proper analysis of the covariance matrix would require a large number of large volume, high resolution simulations, our aim is to show the relative constraining power of the new VDF and the more traditional HMF with dynamical masses.  The covariance matrix is sensitive to changes in cosmological parameters \citep{2009A&A...502..721E} and to properly examine this, one would need a suite of large volume, high resolution simulations in multiple cosmologies spanning the \model{} ranges in consideration.  Such a suite of simulations is not available, so we make the simplifying assumptions that the true covariance matrix does not change significantly over the final $y$+CMB TT likelihood regions shown in Figure \ref{fig:likelihood} and can be approximated by \smash{$\hat{C}$} (Equation \ref{eq:cov}).  Further, we expect that the relative constraining power of the HMF and VDF can be fairly compared even with this simplifying assumption.  The VDF is less sensitive to systematics than the HMF and can utilize upcoming observations to provide further constraints on cosmological parameters.

\section{Conclusion}
\label{sec:conclusion}

The velocity distribution function (VDF) is a novel method for quantifying the abundance of galaxy clusters and constraining cosmological parameters with dynamical cluster observations.  The VDF is a summed PDF of absolute values of galaxy LOS velocities, $|\vlos|$, given by Equation \ref{eq:VDF}.  It is modeled using mock cluster catalogs constructed from the Multidark MDPL1 publicly available $N$-body simulation.  

This new method is compared to a more standard technique for using clusters to constrain cosmological parameters, the halo mass function (HMF).  The HMF compares mock cluster counts as a function of mass to analytic predictions for flat $\Lambda$CDM cosmologies with varying matter density parameter, \OM{}, and amplitude of matter fluctuations, \sig{}.  

The VDF and HMF are used to forecast constraints on \sig{} and \OM{} in the case where true LOS velocities ($\vtrue{}$) and cluster masses ($\Mtot{}$) are known.  Both statistics are also evaluated after introducing reasonable measurement errors in the form of LOS velocity scatter and dynamical cluster masses, and after the introduction of interlopers.  Our main findings can be summarized as follows:

\begin{enumerate}
\item When constructed with perfectly known cluster measurements ($\vtrue{}$ and $\Mtot{}$), the VDF and HMF provide similar constraints on \sig{} and \OM{}.  These constraints can be approximated as a band in the \model{} plane bounded by $\sigma_8\,\Omega_m^\gamma=\mathcal{A}\pm \delta \mathcal{A}_{68\%}$ and with a maximum likelihood ridge running through the center of this band.
\item For an example observation of 200 massive clusters, the VDF {applied to the \new{Idealized} Catalog} constrains the parameter combination \updated{\smash{$\sigma_8\Omega_m^{0.30} = 0.579 \pm 0.011 $}}.  {Applied to the Intermediate Catalog}, the constraint on $\sigma_8\,\Omega_m^\gamma$ is comparable:  \updated{\smash{$\sigma_8\Omega_m^{0.29} = 0.584 \pm 0.011$}}.  {The VDF applied to the Contaminated Catalog constrains the parameter combination  \updated{\smash{$\sigma_8\Omega_m^{0.29} = 0.589 \pm 0.014$}}.}  The VDF maximum likelihood model shows only minor bias to the introduction of {velocity errors, and widens with the introduction of interlopers.}

\item When dynamical mass estimates are used in the HMF, the resulting maximum likelihood \sig{} and \OM{} are strongly biased.  If the scatter in $\Mdyn{}$ is uncorrected, the HMF favors a high \sig{} and low \OM{}.  Additionally, the fiducial model lies well outside of the forecast 2-$\sigma$ constraints.
\end{enumerate}
  
A major source of error in the HMF approach is in connecting cluster observables with halo mass, and in understanding bias and scatter in the observable-mass relationship.  The VDF, however, is less sensitive to such systematics.  Directly using LOS velocities as a cosmological probe mitigates the effects of outliers and minimizes bias.  Further, because simulations of gravity and dynamics are well-understood, the VDF can be explored with $N$-body simulations, in contrast to other cluster-based methods that can be biased by the modeling of gas physics.  Upcoming surveys will probe larger volumes and fainter magnitudes, providing an ideal data set for applying the VDF as a cosmological probe to constrain cosmological parameters.

\acknowledgments{ 
We thank  {our referee for their helpful comments on the manuscript, and also thank Eiichi Egami}, Eiichiro Komatsu, Arthur Kosowsky, Rachel Mandelbaum, Crist{\'o}bal Sif{\'o}n, and David Spergel for their valuable feedback on this
project.
This work is supported in part by DOE DE-SC0011114 grant. 
The CosmoSim database used in this paper is a service by the Leibniz-Institute for Astrophysics Potsdam (AIP).
The MultiDark database was developed in cooperation with the Spanish MultiDark Consolider Project CSD2009-00064.
The Bolshoi and MultiDark simulations have been performed within the Bolshoi project of the University of California High-Performance AstroComputing Center (UC-HiPACC) and were run at the NASA Ames Research Center. The MultiDark-Planck (MDPL) and the BigMD simulation suite have been performed in the Supermuc supercomputer at LRZ using time granted by PRACE.\\ }

\begin{turnpage}

\begin{deluxetable*}{l r r  l r r r r r r r} 

\tabletypesize{\scriptsize}

\tablecaption{Method Summary\label{table:summary}} 
\tablewidth{0pc} 
\tablehead{ 
\colhead{Method} 
&\colhead{Catalog}
&\colhead{$N$} 
&\colhead{Summary}  
& \colhead{$c$\tablenotemark{a}}
& \colhead{$\gamma$\tablenotemark{b}}   
& \colhead{$\mathcal{A}$\tablenotemark{b}}   
& \colhead{$\Delta\sigma_8$\tablenotemark{c}}  
& \colhead{$({\sigma_8})_{68\%}$ \tablenotemark{d}} 
& \colhead{$\Delta\Omega_m$ \tablenotemark{c}} 
& \colhead{${(\Omega_m)}_{68\%}$ \tablenotemark{d}}
}

\startdata 
\\
CMB TT &	 \nodata &	 \nodata 					& TT correlation data for $\ell\leq 2508$		& \nodata & \nodata &\nodata &\nodata & $0.023$ & \nodata  & $0.016$  \\[0.7ex]
\\
\hline
\\
HMF($\Mtot{}$) & \new{Idealized}  & 50 & True cluster mass & \nodata & $ 0.38 $ & $ 0.5246 \pm 0.0196 $ & \nodata & 0.020 & \nodata & 0.015 \\ [0.7ex]
VDF($M$) & \new{Idealized}  & 50 & True $v$, mass sort & \nodata & $ 0.29 $ & $ 0.5857 \pm 0.0259 $ & \nodata & 0.021 & \nodata & 0.016 \\ [0.7ex]
VDF(${\cal R}$) & \new{Idealized}  & 50 & True $v$, richness sort & \nodata & $ 0.27 $ & $ 0.5978 \pm 0.0289 $ & \nodata & 0.020 & \nodata & 0.020 \\ 
\\
HMF($\Mtot{}$) & Intermediate  & 50 & Power law dynamical mass & $ 69.8 \%$ & $ 0.43 $ & $ 0.5068 \pm 0.0154 $ & $ 0.010 $ & 0.020 & $ 0.001 $ & 0.014 \\ [0.7ex]
VDF($M$) & Intermediate  & 50 & $v$ with error, mass sort & $ 1.8 \%$ & $ 0.29 $ & $ 0.5872 \pm 0.0270 $ & $ 0.001 $ & 0.020 & $ -0.001 $ & 0.019 \\ [0.7ex]
VDF(${\cal R}$) & Intermediate  & 50 & $v$ with error, richness sort 	& $ 1.0 \%$ & $ 0.28 $ & $ 0.5976 \pm 0.0288 $ & $ 0.001 $ & 0.020 & $ 0.001 $ & 0.016 \\ 
\\
HMF($\Mtot{}$) & Contaminated  & 50 & Interlopers, power law dynamical mass & $ 80.9 \%$ & $ 0.45 $ & $ 0.5113 \pm 0.0179 $ & $ 0.015 $ & 0.022 & $ 0.002 $ & 0.016 \\ [0.7ex]
VDF($M$) & Contaminated  & 50 & Interlopers, $v$ with error, mass sort & $ 1.8 \%$ & $ 0.3 $ & $ 0.5796 \pm 0.0345 $ & $ 0.002 $ & 0.021 & $ -0.001 $ & 0.021 \\ [0.7ex]
VDF(${\cal R}$) & Contaminated  & 50 & Interlopers, $v$ with error, richness sort & $ 1.4 \%$ & $ 0.29 $ & $ 0.5847 \pm 0.0363 $ & $ 0.001 $ & 0.021 & $ -0.009 $ & 0.042 \\ 
\\
HMF($\Mtot{}$) & \new{Idealized}  & 100 & True cluster mass & \nodata
& $ 0.37 $ & $ 0.5321 \pm 0.0130 $ & \nodata & 0.019 & \nodata & 0.015 \\ [0.7ex]
VDF($M$) & \new{Idealized}  & 100 & True $v$, mass sort & \nodata
& $ 0.28 $ & $ 0.5890 \pm 0.0183 $ & \nodata & 0.019 & \nodata & 0.016 \\ [0.7ex]
VDF(${\cal R}$) & \new{Idealized}  & 100 & True $v$, richness sort & \nodata
& $ 0.27 $ & $ 0.6013 \pm 0.0190 $ & \nodata & 0.018 & \nodata & 0.016 \\ 
\\
HMF($\Mtot{}$) & Intermediate  & 100 & Power law dynamical mass & $ 87.9 \%$ & $ 0.42 $ & $ 0.5137 \pm 0.0115 $ & $ 0.014 $ & 0.020 & $ -0.002 $ & 0.014 \\ [0.7ex]
VDF($M$) & Intermediate  & 100 & $v$ with error, mass sort & $ 3.5 \%$ & $ 0.29 $ & $ 0.5893 \pm 0.0186 $ & $ 0.003 $ & 0.019 & $ -0.001 $ & 0.016 \\ [0.7ex]
VDF(${\cal R}$) & Intermediate  & 100 & $v$ with error, richness sort 	& $ 1.9 \%$ & $ 0.27 $ & $ 0.6016 \pm 0.0191 $ & $ 0.001 $ & 0.019 & $ 0.001 $ & 0.016 \\ 
\\
HMF($\Mtot{}$) & Contaminated  & 100 & Interlopers, power law dynamical mass & $ 94.7 \%$ & $ 0.43 $ & $ 0.5221 \pm 0.0134 $ & $ 0.021 $ & 0.021 & $ 0.001 $ & 0.014 \\ [0.7ex]
VDF($M$) & Contaminated  & 100 & Interlopers, $v$ with error, mass sort & $ 3.0 \%$ & $ 0.29 $ & $ 0.5858 \pm 0.0237 $ & $ 0.002 $ & 0.020 & $ -0.001 $ & 0.016 \\ [0.7ex]
VDF(${\cal R}$) & Contaminated  & 100 & Interlopers, $v$ with error, richness sort & $ 2.5 \%$ & $ 0.29 $ & $ 0.5847 \pm 0.0248 $ & $ 0.002 $ & 0.020 & $ 0.001 $ & 0.016 \\ 
\\
HMF($\Mtot{}$) & \new{Idealized}  & 200 & True cluster mass & \nodata
& $ 0.37 $ & $ 0.5314 \pm 0.0082 $ & \nodata & 0.017 & \nodata & 0.014 \\ [0.7ex]
VDF($M$) & \new{Idealized}  & 200 & True $v$, mass sort & \nodata
& $ 0.30 $ & $ 0.5785 \pm 0.0106 $ & \nodata & 0.017 & \nodata & 0.015 \\ [0.7ex]
VDF(${\cal R}$) & \new{Idealized}  & 200 & True $v$, richness sort & \nodata
& $ 0.28 $ & $ 0.5887 \pm 0.0112 $ & \nodata & 0.017 & \nodata & 0.016 \\ 
\\
HMF($\Mtot{}$) & Intermediate  & 200 & Power law dynamical mass & $ 99.3 \%$ & $ 0.42 $ & $ 0.5103 \pm 0.0094 $ & $ 0.023 $ & 0.018 & $ -0.006 $ & 0.014 \\ [0.7ex]
VDF($M$) & Intermediate  & 200 & $v$ with error, mass sort & $ 8.7 \%$ & $ 0.29 $ & $ 0.5841 \pm 0.0107 $ & $ 0.004 $ & 0.017 & $ -0.001 $ & 0.015 \\ [0.7ex]
VDF(${\cal R}$) & Intermediate  & 200 & $v$ with error, richness sort 	& $ 4.9 \%$ & $ 0.28 $ & $ 0.5938 \pm 0.0109 $ & $ 0.003 $ & 0.017 & $ -0.001 $ & 0.016 \\ 
\\
HMF($\Mtot{}$) & Contaminated  & 200 & Interlopers, power law dynamical mass & $ 99.9 \%$ & $ 0.44 $ & $ 0.5148 \pm 0.0107 $ & $ 0.033 $ & 0.018 & $ -0.002 $ & 0.013 \\ [0.7ex]
VDF($M$) & Contaminated  & 200 & Interlopers, $v$ with error, mass sort & $ 7.6 \%$ & $ 0.29 $ & $ 0.5892 \pm 0.0137 $ & $ 0.004 $ & 0.018 & $ -0.001 $ & 0.015 \\ [0.7ex]
VDF(${\cal R}$) & Contaminated  & 200 & Interlopers, $v$ with error, richness sort & $ 7.0 \%$ & $ 0.29 $ & $ 0.5854 \pm 0.0143 $ & $ 0.003 $ & 0.018 & $ 0.001 $ & 0.016 \\ 
\\

\enddata 

\tablenotetext{a}{Smallest contour containing the fiducial model, see Equation \ref{eq:contour}.}
\tablenotetext{b}{Parameterization of band in \model{} plane; see Equation \ref{eq:band}.}
\tablenotetext{c}{Maximum likelihood offset after combining with CMB TT constraints; see Equation \ref{eq:DeltaCMB}.}
\tablenotetext{d}{68\% confidence after combining with CMB TT constraints.}

\end{deluxetable*} 
\end{turnpage}

\end{document}